\documentclass[preprint,12pt,authoryear]{elsarticle}
\usepackage{tikz}
\usetikzlibrary{positioning,arrows.meta,fit,backgrounds}
\usepackage{amsmath}
\usepackage{graphicx}
\usepackage{enumerate}
\usepackage[round,authoryear]{natbib}
\usepackage{amsfonts,amssymb,epsfig,verbatim,bm,latexsym,amsbsy}
\usepackage{tikz}
\usepackage{float}
\usetikzlibrary{shapes,arrows}
\usepackage{rotating}
\usepackage{algorithm}
\usepackage{algorithmic}
\usepackage{url}
\usepackage{amsthm}
\usepackage{booktabs}
\usepackage{multirow}
\usepackage{longtable}
\usepackage{array}
\usepackage{pdflscape}
\usepackage{ragged2e}
\usepackage{array}

\usepackage[top=2cm, bottom=1.5cm, left=2cm, right=2cm]{geometry}
\usepackage{epstopdf}
\usepackage{multicol}
\usepackage{multirow}

\journal{Statistical Methods and Applications}

\begin{document}

\begin{frontmatter}


\title{Multicollinearity diagnosis in generalized linear models: A systematic review and empirical reassessment}


\author{Catalina B. Garc\'ia-Garc\'ia \corref{cor1}}
\author{Jorge L. Bazán Guzmán}
\cortext[cor1]{Corresponding author}
\author{Rom\'an Salmer\'on-G\'omez}


\begin{abstract}
Ridge-type estimators are widely used to address multicollinearity in generalized linear models (GLMs), yet their empirical justification depends heavily on how multicollinearity is diagnosed. Following systematic review guidelines, we conduct a citation-based review of 120 real-data applications to examine current diagnostic practices. The findings reveal a lack of consensus across studies: approaches vary regarding data transformation prior to diagnosis, applying transformations to either the design matrix or the weighted information matrix, while others omit transformation entirely. Nevertheless, the condition number emerges as the most widely used diagnostic. We review existing methodologies for computing the condition number in GLMs and recommend the optimal approach, as it uniquely attains its minimum value under orthogonality while retaining the intercept. Finally, these formulations are implemented in R and used to replicate five real-data applications from the literature.
\end{abstract}



\end{frontmatter}

\section{Introduction}
Generalized linear models (GLMs) provide a flexible framework for modelling a wide range of response variables belonging to the exponential family and have become one of the most widely used modelling approaches in applied statistics. By extending the classical linear model through the specification of an appropriate probability distribution and link function, GLMs allow the relationship between a response variable and a set of explanatory variables to be modelled within a unified framework. Their flexibility has led to extensive applications in fields such as epidemiology, medicine, ecology, economics, and the social sciences.

As in linear regression, strong dependencies among explanatory variables can represent an important problem in GLMs. Multicollinearity happens when two or more independent variables (including the intercept) are highly correlated with one another, resulting in an ill-conditioned estimation problem. Although multicollinearity does not necessarily affect the predictive ability of a model, it can substantially increase the variance of parameter estimates, produce unstable coefficient estimates, and make the magnitude and direction of estimated effects difficult to interpret, \cite{willan1978meaningful}. These problems are particularly relevant in GLMs because parameter estimation is based on a model-dependent information matrix that incorporates not only the design matrix but also the weights determined by the fitted model.

\cite{mackinnon1989collinearity} addressed the problem of multicollinearity in GLMs and established an early methodological basis for its treatment through biased estimation. This work highlighted the potential of ridge estimation for reducing the adverse effects of ill-conditioning and obtaining more stable parameter estimates. Similarly, \cite{weissfeld1991multicollinearity} expanded multicollinearity diagnostics for discrete GLM response models by adapting the Belsley--Kuh--Welsch framework to evaluate the weighted expected information matrix through condition indices and variance decomposition proportions. Subsequently, \cite{segerstedt1992ordinary} developed ridge estimation further within the GLM framework, providing an important foundation for the subsequent development of ridge-type estimators in GLMs.

The contributions of \cite{mackinnon1989collinearity}, \cite{weissfeld1991multicollinearity} and \cite{segerstedt1992ordinary} have inspired substantial and growing bodies of research. According to Web of Science database, 57, 39 and 130 studies\footnote{Figures from Web of Science database, accessed on August 11, 2026.}, respectively, have cited these three contributions. More than 75\% of the studies citing \cite{mackinnon1989collinearity}, more than 66\% of those citing \cite{weissfeld1991multicollinearity} and more than 80\% of those citing \cite{segerstedt1992ordinary} were published since 2020. Collectively, these citing studies have received 759, 573 and 1636 citations, respectively, of which more than 89\%, 78\% and 91\% have been accumulated since 2020. These figures highlight the substantial and particularly recent expansion of research stemming from these three foundational contributions.

Despite this substantial development, most contributions have focused on the estimation problem itself, particularly on the construction and comparison of ridge and alternative shrinkage estimators or on the selection of the ridge penalty parameter. Comparatively little attention has been devoted to the prior question of how multicollinearity is diagnosed in real-data applications and whether the diagnostic procedures used provide adequate empirical justification for the subsequent use of ridge estimation.  Among the studies citing  \cite{segerstedt1992ordinary}, \cite{review42} is, to our knowledge, the contribution that focuses most explicitly on the diagnosis of multicollinearity. The author proposes new criteria for both maximum likelihood and ridge estimators and investigate their properties for different values of the ridge parameter $k$. They also propose an extension of the condition number for use after ridge estimation and suggest a threshold of 100. Their numerical illustrations consider Poisson, binomial, and gamma models and examine the effects of centering and scaling the information matrix. This contribution builds upon the earlier work of \cite{weissfeld1991multicollinearity}, which emphasized the importance of working with unit length transformed data when condition-based diagnostics are used. It should be noted, however, that in the design matrix $\mathbf{X}$ considered by \cite{review42} the intercept  is not taken into account in the resulting multicollinearity diagnostics.

This omission is not unique to \cite{review42}. Across the broader set of diagnostic measures used in the reviewed literature (correlation-matrix, variance inflation factors, condition indices, condition numbers and eigenvalue-based diagnostics), the treatment of the intercept is rarely stated explicitly and, when reported, varies considerably across studies. This lack of clarity is not merely a reporting issue, as the treatment of the intercept determines which type of multicollinearity a given diagnostic can detect. Multicollinearity is traditionally classified into two types: essential multicollinearity, which arises from strong relationships among the explanatory variables, excluding the intercept, and non-essential multicollinearity, which results from a strong relationship between the intercept and at least one explanatory variable (\cite{Marquardt1980,salmeron2020diagnosis}). However, neither of the two most commonly used diagnostics in multiple linear regression (the variance inflation factor (VIF) and pairwise correlations) is able to detect non-essential multicollinearity. Similarly, the condition number (CN) cannot detect non-essential multicollinearity when it is computed from a matrix that excludes the intercept.

The rapid expansion of empirical applications of ridge-type estimators therefore raises an important question: to what extent is multicollinearity appropriately diagnosed in real-data applications in GLMs?  The existing literature has largely examined multicollinearity diagnostics, ridge estimation, and GLMs as separate methodological topics. For example, \cite{alin2010multicollinearity} reviews common multicollinearity diagnostics, including correlations, variance inflation factors, and condition indices, whereas \cite{mcdonald2009ridge,mcdonald2010tracing} discuss ridge regression and ridge traces as approaches to mitigating multicollinearity. These contributions are mainly framed within the classical linear-model setting. Conversely, \cite{neuhaus2011generalized} reviews the GLM framework without focusing specifically on multicollinearity diagnosis. Despite the widespread adoption of ridge estimation in GLMs, the systematic link between diagnostic methods for multicollinearity and the empirical justification for subsequent estimation choices remains poorly understood and largely unexamined.

To bridge this gap, this study pursues three key objectives. First, we systematically investigate how multicollinearity is diagnosed in GLMs in empirical practice. Second, we identify the most appropriate
mathematical formulation for computing the most widely used diagnostic. Third, we provide an R implementation, replicate multiple literature datasets, and outline concrete practical guidelines. Following the PRISMA 2020 framework, our systematic review maps empirical applications citing three seminal index papers: \cite{mackinnon1989collinearity}, \cite{weissfeld1991multicollinearity} and \cite{segerstedt1992ordinary}. Building on these insights, we resolve existing computational inconsistencies and present real-data replications to guide applied researchers.

The remainder of the paper is organized as follows.  Section \ref{glm_structure} reviews the underlying structure of GLMs then presents the systematic and the main methodological aspects of multicollinearity diagnosis.
Section \ref{review} presents the systematic, citation-based literature review of studies citing  \cite{mackinnon1989collinearity}, \cite{weissfeld1991multicollinearity} and \cite{segerstedt1992ordinary}, including the search strategy, eligibility criteria and study selection, including different subsections to analyze the methodological issues identified in the reviewed literature: studies that do not report any assessment of multicollinearity, studies relying exclusively on partial diagnostics, the use of the original information matrix $\mathbf{X}'\mathbf{X}$ rather than the appropriate GLM information matrix, misleading with the condition number and the condition index, the absence of the recommended unit-length transformation and the lack of distinction between essential and non-essential multicollinearity. Section \ref{methods} reviews the methodological proposals for computing the condition number in GLMs and compares the approaches of \cite{mackinnon1989collinearity} and \cite{weissfeld1991multicollinearity}. Section \ref{package} replicates five real-data applications from the literature. Finally, Section \ref{conclusions}  summarizes the main conclusions and outlines directions for future research.

\section{Review of underlying structure of GLMs} \label{glm_structure}

Consider a sample of independent observations \((y_i,\mathbf{x}_i)\), \(i=1,\ldots,n\), where \(y_i\) is the response and \(\mathbf{x}_i\in\mathbb{R}^p\) contains the explanatory variables. A GLM has three components:
\begin{itemize}
\item Random component. Conditional on \(\mathbf{x}_i\), the response \(Y_i\) has an exponential-family density or probability mass function
\[ f(y_i;\theta_i,\phi) = \exp\left\{ \frac{y_i\theta_i-b(\theta_i)}{a(\phi)} +c(y_i,\phi) \right\}, \]
for some specific functions $a(.), b(.), c(.)$ where \(\theta_i\) is the natural parameter and \(\phi\) is a dispersion parameter. If $\phi$ is known, this is an exponential family model with canonical parameter $\theta$.
Consequently,
\[
\mu_i=\mathbb{E}(Y_i)=\frac{db(\theta_i)}{d\theta_i},
\qquad
\operatorname{Var}(Y_i)=a(\phi)\frac{d^2b(\theta_i)}{d\theta_i^2}.
\]
\item Linear predictor.
\[ \eta_i=\mathbf{x}'_i\boldsymbol{\beta}, \]where \(\boldsymbol{\beta}\) is an unknown vector of regression coefficients.
\item Link function. A known, invertible function \(g\) connects the conditional mean to the linear predictor:
\[ g(\mu_i)=\mathbf{x}'_i\boldsymbol{\beta}. \]The link is canonical when \(\eta_i=\theta_i\).
\end{itemize}

Note that in a GLM, the variance is generally not constant, instead, it is modeled as a function of the mean. By following the notation of \cite{mccullagh2019generalized}, a GLM assumes that $Y_i$ has variance $\mathrm{V}(Y_i)=\phi V(\mu_i)$, with mean $\mu_i=g^{-1}(\eta_i)$ and linear predictor $\eta_i = X_i'\mathbf{\beta}$, where $g(\cdot)$ is a link function. In the iterative reweighted least squares (IRLS) representation (\cite{nelder1972generalized}), the maximum likelihood estimator is obtained iteratively as:
$$
\hat{\boldsymbol{\beta}}^{(m+1)} = (\mathbf{X}' \mathbf{W}^{(m)} \mathbf{X})^{-1} \mathbf{X}' \mathbf{W}^{(m)} \mathbf{z}^{(m)},
$$
where $\mathbf{W}^{(m)}=\mathrm{diag}(w_{ii})$ is the $n \times n$ diagonal matrix with weights $w_{ii}=(1/v_i)(d\mu_i/d\eta_i)^2$ and $\mathbf{z}^{(m)}$ is the $n \times 1$ working-response vector with elements:
$$z_i^{(m)}=\sum_{j=1}^px_{ij}\hat{\beta}_j^{(m)}+(y_i-\hat{\mu_i})\frac{d\eta_i}{d\mu_i},\quad i=1,...,n,$$
$v_i$, $\hat{\mu}_i$ and $d\mu_i/d\eta_i$ are treated at $\hat{\boldsymbol{\beta}}^{(m)}$ and $m$ denotes the iteration step.  This procedure is repeated until convergence. Common convergence criteria include convergence of the parameter estimates, the log-likelihood, and the Hessian matrix, see \cite{review42}. Upon convergence, the maximum likelihood estimator is given by:
$$\hat{\boldsymbol{\beta}}_{ML}=(\mathbf{X}'\hat{\mathbf{W}}\mathbf{X})^{-1}\mathbf{X}'\hat{\mathbf{W}}\hat{\mathbf{z}}$$

In this context, by using the matrix of Fisher information for $\mathbf{\beta}$:
$$I(\boldsymbol{\beta})=\mathbf{X}'\mathbf{\hat{W}}\mathbf{X}$$
the covariance matrix of the maximum likelihood (ML) estimator is given by $var(\boldsymbol{\beta})=\phi (\mathbf{X}'\hat{\mathbf{W}}\mathbf{X})^{-1}$.

Note that both $w_{ii}$ and $z_i$ depend on the distribution and the link function. When the response follows a normal distribution, $Y_i \sim \mathcal{N}(\mu_i, \sigma^2)$, the canonical link is the identity,
$g(\mu)=\mu$, and the variance function is constant, $V(\mu_i)=1$. Consequently, the derivative $d\mu_i/d\eta_i=1$, and the
weights reduce to $w_{ii} = 1/\sigma^2$, so the weight matrix is proportional to the identity. The working response simplifies to
$z_i = y_i$, recovering the classical ordinary least squares formulation. This simplification generally does not occur for other distributions, where the variance depends on the mean and the weights vary across observations. For such cases, the IRLS formulation adjusts both the weight matrix $\mathbf{W}$ and the working response $\mathbf{z}$ at each iteration, reflecting the heteroscedastic
nature of the data. Appendix A summarizes different distributions and function links.

It is evident that, when the matrix $\mathbf{X}'\mathbf{X}$ is ill-conditioned, for some weight matrix $\hat{\mathbf{W}}$, the matrix $\mathbf{X}'\hat{\mathbf{W}}\mathbf{X}$ can be also ill-conditioned,
and multicollinearity can exist in GLMs. This can have important practical consequences, as an ill-conditioned information matrix may lead to unstable parameter estimates, inflated standard errors, and increased sensitivity of the estimated coefficients to small changes in the data or in the model specification. These effects can complicate the interpretation of individual regression coefficients and affect statistical inference, even when the model as a whole provides an adequate fit.

In this context, \cite{mackinnon1989collinearity} specifically investigated multicollinearity in GLMs, showing that its effects can be related to the conditioning of the estimated information matrix and proposing diagnostic criteria for assessing the degree of multicollinearity in GLMs. Complementarily, \cite{weissfeld1991multicollinearity} extended these multicollinearity diagnostics to discrete response GLMs by adapting the Belsley--Kuh--Welsch framework to evaluate the weighted expected information matrix through condition indices and variance decomposition proportions. \cite{segerstedt1992ordinary} subsequently showed that the effects of an ill-conditioned design matrix on the parameter estimates in GLMs are analogous to those observed in linear regression models, and proposed a generalized ridge estimator to mitigate the resulting instability.

In practice, several diagnostics and rules of thumb have been proposed in the literature to analyze multicollinearity in GLMs, largely building on the three seminal works cited above, frequently adapting criteria originally developed for linear regression. Despite the critical importance of this issue, no single universally accepted diagnostic procedure exists for GLMs. When multicollinearity is detected, ridge regression is among the most widely applied methodologies to estimate parameters under collinearity. However, employing ridge estimation as a remedy should be preceded by an adequate diagnosis to verify whether the reduction in variance justifies the introduced estimation bias. To evaluate how multicollinearity is diagnosed in both methodological developments and empirical applications of GLMs, we conduct a systematic literature review centered on the foundational frameworks of \cite{mackinnon1989collinearity}, \cite{weissfeld1991multicollinearity} and \cite{segerstedt1992ordinary}. The review design, main findings, and discussion are presented in Section 3.

\section{Systematic literature review}\label{review}

A systematic literature review was conducted to assess how multicollinearity is evaluated in real-data applications of ridge and ridge-type estimators for GLMs. The review followed the PRISMA 2020 guidelines to ensure a transparent and reproducible process of study identification, selection, data extraction, and evidence synthesis (\cite{page2021prisma}; \cite{rethlefsen2021prisma}). A citation-based search strategy was adopted using three index studies that established important foundations for the treatment of multicollinearity and ridge estimation in the GLM framework, namely \cite{mackinnon1989collinearity}, \cite{weissfeld1991multicollinearity} and \cite{segerstedt1992ordinary}. The review specifically focused on the diagnostic procedures employed, the thresholds adopted, the matrices on which the diagnostics were calculated, the treatment of variable scaling, and the distinction between essential and non-essential multicollinearity.

The Web of Science Core Collection (WoS) was used as the bibliographic source. Cited-reference searches were performed for the index studies on August 11, 2026, identifying 57 records citing \cite{mackinnon1989collinearity} , 39 records citing \cite{weissfeld1991multicollinearity} and 130 records citing \cite{segerstedt1992ordinary}. The three searches yielded 226 records in total, of which 47 were common to all citation sets. After accounting for these overlapping records, 179 unique records constituted the initial pool for screening and eligibility assessment. Studies were eligible if they (i) explicitly cited at least one of the three index studies; (ii) used, developed, compared, or evaluated a ridge or ridge-type estimator within the GLM framework; and (iii) included an application to real data, regardless of whether theoretical or simulation results were also reported. Studies that merely cited an index study without a substantive methodological contribution, exclusively theoretical or simulation-based studies without real-data applications, studies not involving GLMs or ridge-type estimation, and documents not meeting the predefined publication criteria were excluded.

\begin{figure}[htbp]
\centering
\begin{tikzpicture}[
    node distance=7mm and 7mm,
    source/.style={
        rectangle,
        rounded corners,
        draw=black,
        align=center,
        text width=3.8cm,
        minimum height=1.1cm,
        font=\small
    },
    process/.style={
        rectangle,
        rounded corners,
        draw=black,
        align=center,
        text width=7.7cm,
        minimum height=1cm,
        font=\small
    },
    exclusion/.style={
        rectangle,
        rounded corners,
        draw=black,
        align=left,
        text width=4.5cm,
        minimum height=2.0cm,
        font=\small,
        inner sep=6pt
    },
    arrow/.style={-{Stealth[length=2.5mm]}, thick}
]

\node[source] (mac) {
    \textbf{\cite{mackinnon1989collinearity}}\\
    57 citing records
};

\node[source, right=of mac] (weis) {
    \textbf{\cite{weissfeld1991multicollinearity}}\\
    39 citing records
};

\node[source, right=of weis] (seg) {
    \textbf{\cite{segerstedt1992ordinary}}\\
    130 citing records
};

\node[process, below=12mm of weis] (identified) {
    \textbf{Records identified through the three source articles}\\
    $n=226$
};

\node[process, below=of identified] (duplicates) {
    \textbf{Duplicate records removed}\\
    $n=47$
};

\node[process, below=of duplicates] (screened) {
    \textbf{Records screened for eligibility}\\
    $n=179$
};

\node[process, below=of screened] (excluded) {
    \textbf{Records excluded}\\
    $n=59$
};

\node[exclusion, right=12mm of excluded] (reasons) {
    \textbf{Reasons for exclusion}\\[2pt]
    -- Outside the scope of the review\\
    -- No relevant GLM ridge-type application\\
    -- No relevant real-data analysis\\
    -- Other eligibility criteria
};

\node[process, below=of excluded] (included) {
    \textbf{Studies included in the review}\\
    $n=120$
};

\draw[arrow] (mac.south) |- (identified.west);
\draw[arrow] (weis.south) -- (identified.north);
\draw[arrow] (seg.south) |- (identified.east);

\draw[arrow] (identified) -- (duplicates);
\draw[arrow] (duplicates) -- (screened);
\draw[arrow] (screened) -- (excluded);
\draw[arrow] (excluded) -- (included);

\draw[arrow] (excluded.east) -- (reasons.west);

\end{tikzpicture}
\caption{PRISMA-style flow diagram of the citation-based systematic review.}
\label{fig:prisma}
\end{figure}
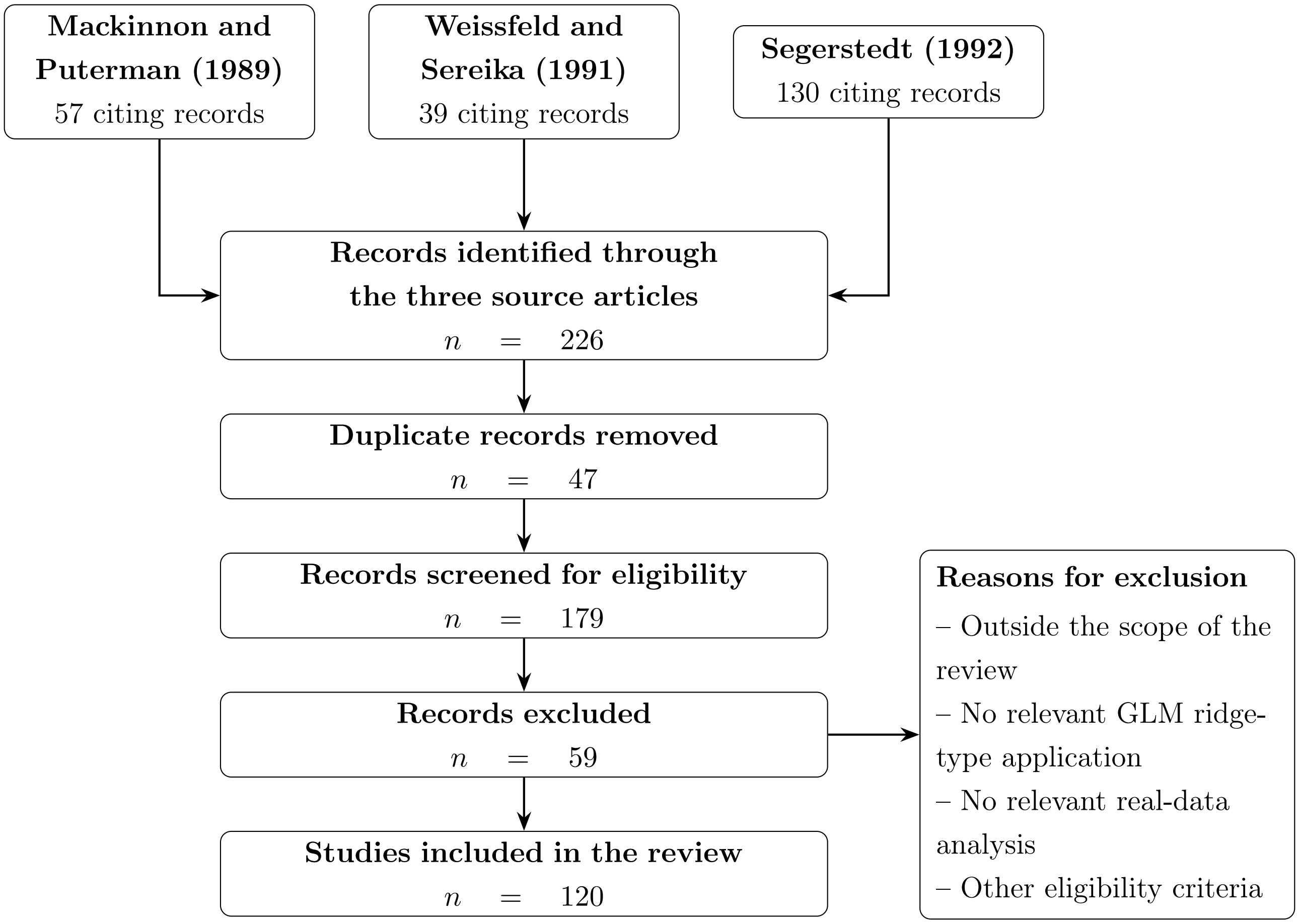

The screening process involved an initial assessment of titles and abstracts, followed by a full-text assessment of potentially eligible records. Reasons for exclusion at the full-text stage were systematically documented. Of the 179 unique records screened, 59 were excluded, leaving 120 studies in the final review. Among the included studies, 76 cited only \cite{segerstedt1992ordinary}, 10 cited only \cite{mackinnon1989collinearity}, and 6 cited only \cite{weissfeld1991multicollinearity}; the remaining 28 cited at least two of the three index studies. The study selection process is illustrated in the PRISMA 2020 flow diagram shown in Figure \ref{fig:prisma}.




Importantly, absence of reported information was distinguished from evidence that a procedure was not performed. When an article did not report whether multicollinearity had been assessed, this was coded as ``not reported'' rather than interpreted as evidence that no diagnostic procedure had been conducted. In the following subsections, we analyze the different methodological issues identified in the reviewed literature: works that do not report any measure related to the diagnosis of multicollinearity; works that report only the correlation matrix or variance inflation factors; works that diagnose multicollinearity from the unweighted matrix $\mathbf{X}'\mathbf{X}$; works that do not adequately transform the data, or do not report whether any transformation was applied; works that report the condition index as a single value; and works that do not report a threshold against which to compare the condition number, or that compare it against a misleading threshold.

\subsection{Disregarding Multicollinearity}
Eight of the 120 included studies (6.67\%) applied a ridge-type estimator to real data without reporting any diagnostic assessment of multicollinearity prior to estimation \citep{review87,review102,review111,review138,review110,review128,review120,review119}. These studies span a range of GLM families (binary logistic, multinomial ordinal, and gamma/gamma-Pareto regression) and, consistent with the coding convention adopted in this review, the absence of a reported diagnostic is treated strictly as ``not reported'' rather than as evidence that no such assessment was ever performed. Two of the eight, \cite{review120} and \cite{review119}, share the same authors, real dataset, and Gamma-Pareto ridge model, and are primarily concerned with influence diagnostics and residual behavior rather than with multicollinearity per se, which may explain why a multicollinearity-specific diagnostic was not foregrounded. Two others, \cite{review111} and \cite{review128}, are situated in high- or ultra-high-dimensional settings, where standard multicollinearity diagnostics such as variance inflation factor (VIF) or the condition number (CN) are less routinely applied and may be considered less tractable than in the low-dimensional applications that dominate the rest of the corpus. The remaining four, \cite{review87}, \cite{review102}, \cite{review138} and \cite{review110}, cover binary, ordinal, zero inflated poisson and gamma distributed responses respectively and share no obvious common feature beyond applying a ridge-type estimator to a real dataset without a stated multicollinearity check. Taken together, these eight cases illustrate that even within a literature explicitly motivated by multicollinearity, the empirical justification for applying a ridge-type remedy is not universally documented.

\subsection{Partial diagnostic: limitations of correlation-based and VIF diagnostics}

A second group of studies relies exclusively on pairwise correlation coefficients and/or variance inflation factors (VIFs) to assess multicollinearity without complementing them with a condition index or condition number \citep{review4,review22,review52,review28,review85,review93,review21,review59,review58,review78,review92}. These 11 (9.17\%) studies span a heterogeneous set of GLM families (multinomial logit, gamma, inverse Gaussian, zero-inflated negative binomial, Cox/Weibull survival, and binomial logistic models) indicating that reliance on correlation and/or VIF alone is not confined to a particular response distribution.

Although these measures are widely used in applied research, their use as the sole diagnostic can provide an incomplete assessment of the underlying multicollinearity structure. Pairwise correlations evaluate linear association between pairs of predictors and therefore do not constitute a global measure of multicollinearity. In particular, substantial multivariate dependencies may exist even when all pairwise correlations are relatively moderate. Similarly, VIFs quantify the extent to which the variance of an individual regression coefficient is inflated as a result of its linear relationship with the remaining predictors, but they do not directly characterize the overall dimensional structure of the design matrix. Moreover, both diagnostics may be problematic in applications involving categorical or binary predictors, depending on how the variables are coded and on the structure of the model, for example, \cite{review78} calculates the VIF and correlation matrix including dichotomic variables.

More importantly, correlation-based and VIF diagnostics do not capture non-essential multicollinearity, \citep{salmeron2020diagnosis,salmeron2022multicoll}. Consequently, an assessment based exclusively on pairwise correlations or VIFs may fail to provide a complete picture of the sources and structure of ill-conditioning in the design matrix. This issue is particularly relevant in the paper presented by \cite{review59}, which includes interaction terms among their explanatory variables, due to the potential for these terms to induce linear or near-linear dependencies that are not captured by pairwise correlations or VIFs. 

\subsection{Partial diagnostic: Analyzing multicollinearity from $\mathbf{X}'\mathbf{X}$}

In contrast to the majority pattern found across the rest of the review, five of the included studies compute their multicollinearity diagnostic on the unweighted matrix $\mathbf{X}'\mathbf{X}$ rather than on the IRLS-weighted $\mathbf{X}'\hat{\mathbf{W}}\mathbf{X}$. \cite{review3}, \cite{review6} and \cite{review126} (in the first of its two real-data applications) all fit Poisson-family regressions to the same aircraft damage data and report the eigenvalues of $\mathbf{X}'\mathbf{X}$ excluding the intercept and obtain the same value for the condition number (219.3654) without reweighting it by the model's own fitted values; neither reports centering or scaling the covariates for this calculation, and \cite{review3} seems to reserve the unit-length-scaled, weighted $\mathbf{X}'\hat{\mathbf{W}}\mathbf{X}$ approach for its simulation study only, not for the real-data application. \cite{review126}'s second application, a negative binomial model fitted to the pedestrian-fatalities data, follows the same unweighted approach. \cite{review63}, fitting a Bell regression to the bioChemists publication-count data, likewise reports a condition number (984.1666) computed from the unweighted $\mathbf{X}'\mathbf{X}$, again without stating whether the covariates were centered or scaled beforehand. \cite{review117} (discrete Weibull regression, mine fracture data) is the least transparent of the five: it computes the condition number from the unweighted $\mathbf{X}'\mathbf{X}$ with no covariate transformation, and this can only be verified from the appendix R code, since no numerical value is reported in the manuscript itself.

In addition, of the reviewed articles, 36 report a numerical multicollinearity diagnostic (a condition index, condition number, or VIF) in their real-data application without explicitly stating, at the point of computation, which matrix —$\mathbf{X}'\mathbf{X}$ or the IRLS-weighted $\mathbf{X}'\hat{\mathbf{W}}\mathbf{X}$— that diagnostic was calculated on. For the majority of these, 29 articles, the ambiguity is total: neither the applied section nor the surrounding theory commits to a specific matrix. This is the case for \cite{review5}, \cite{review11}, \cite{review116}, \cite{review18}, \cite{review26}, \cite{review38}, \cite{review50}, \cite{review94}, \cite{review31}, \cite{review77},  \cite{review65}, \cite{review84}, \cite{review88}, \cite{review89}, \cite{review90}, \cite{review100}, \cite{review101}, \cite{review105},\cite{review124}, \cite{review125},\cite{review157}, \cite{review49}, \cite{review36}, \cite{review80}, \cite{review99}, \cite{review112}, \cite{review115}, \cite{review17} and \cite{review_w16}. For a further 4 articles (\cite{review16}, \cite{review72}, \cite{review62} and \cite{review71}) the matrix can be inferred with reasonable confidence even though it is not restated in the application section, because the theoretical development defines and uses exclusively the weighted matrix $\mathbf{X}'\hat{\mathbf{W}}\mathbf{X}$ throughout, with the proposed ridge- or Liu-type estimator itself built directly from its eigen decomposition. The remaining 3 articles are inconsistent across their own multiple real-data applications. \cite{review113} names the matrix explicitly only for its second example (reported as $\mathbf{X}'\hat{\mathbf{W}}\mathbf{X}$), while its first example reports a bare condition number with no matrix specified. \cite{review12} names the matrix explicitly only for its third example (reported as $\mathbf{X}'\hat{\mathbf{W}}\mathbf{X}$). \cite{review103} does not name the matrix in its own text at all, but shares the same real-data example and numerical results as \cite{review57}, which does report the diagnostic explicitly on $\mathbf{X}'\hat{\mathbf{W}}\mathbf{X}$, suggesting, though not confirming, that the same matrix underlies both.

\subsection{Inconsistencies in relation to data transformation}

There are 60 articles that explicitly state that they diagnose multicollinearity on the IRLS-weighted matrix $\mathbf{X}'\hat{\mathbf{W}}\mathbf{X}$. Eight of them, however, do so excluding the intercept term: \cite{review8}, \cite{review33}, \cite{review25}, \cite{review108}, \cite{review81}  and \cite{review123} do not provide explicit information on what transformation, if any, was applied to the data before the diagnostic was computed, \cite{review42} excluded the intercept and applied centering and unit length transformation and \cite{review37} stated explicitly that ``The COMP regression model is fitted without any standardization and without intercept.'' Of the remaining 52, 30 do not explicitly state that any transformation was applied to the data: \cite{review14}, \cite{review10}, \cite{review13}, \cite{review147}, \cite{review15}, \cite{review27}, \cite{review109}, \cite{review34}, \cite{review51}, \cite{review129}, \cite{review150}, \cite{review45}, \cite{review47}, \cite{review60}, \cite{review154}, \cite{review9}, \cite{review76}, \cite{review158}, \cite{review23}, \cite{review44}, \cite{review57}, \cite{review61}, \cite{review73}, \cite{review98}, \cite{review127}, \cite{review104}, \cite{review106}, \cite{review_w6}, \cite{review151} and
\cite{review_w28}. For 5 further articles (\cite{review2}, \cite{review55}, \cite{review118}, \cite{review24} and \cite{review97}) no transformation is reported for the real-data application itself, but it can be inferred either from the simulation section of the same paper or from a related, previously cited study using the same data, that the covariates may have been transformed. The remaining 17 explicitly state that a transformation was applied to the data of the real-data application, but they differ both in what is transformed and in whether the intercept is retained in the diagnosed matrix. Five apply unit-length scaling without centering (\cite{review133}, \cite{review_w34}, \cite{review40}, \cite{review83}, \cite{review114}); only \cite{review133} and \cite{review_w34} retain the intercept, whereas the other three compute the diagnostic on a matrix without it, and \cite{review40} applies the transformation in only one of its two applications. Seven center and standardize the covariates so that $X'X$ is in correlation form (\cite{review74}, \cite{review19}, \cite{review155}, \cite{review137}, \cite{review35}, \cite{review_w20}, \cite{ertan2024identifying}): the first five fit the model and compute the diagnostic without an intercept, while \cite{review_w20} and \cite{ertan2024identifying} add a column of ones after centering, which makes the intercept orthogonal to the regressors by construction. Three scale the covariates and then, following \cite{review42}, center and scale the columns of $\hat W^{1/2}X$ before computing the diagnostic, which again removes the intercept (\cite{review39}, \cite{review66}, \cite{review121}). Finally, \cite{review69} and \cite{review54} state that the covariates were standardized without specifying the
scheme. Note that most of the works do not report whether a transformation was applied, and when it is reported is not always explicitly explained.

\subsection{Misleading in relation to the concept of Condition Index}
From the systematic review, we found a recurring terminological inconsistency rather than a substantive computational one. Strictly speaking, the condition index is not a single number: for a design matrix with eigenvalues
$\lambda_1 \geq \lambda_2 \geq \lambda_3 \geq \cdots \geq \lambda_p$, each eigenvalue defines its own condition index $CI_i=\sqrt{\frac{\lambda_{max}}{\lambda_i}}$ with $i=1,...,p$, so a full multicollinearity diagnosis in the sense of \cite{Belsley1980} yields a vector of $p$ condition indices, one per dimension, not a single scalar. What is properly a single number is the condition number, conventionally taken as the largest of these condition indices, $CN=\sqrt{\frac{\lambda_{max}}{\lambda_{min}}}$, that is, the square root of the ratio between the maximum and the minimum eigenvalue. In practice, however, the large majority of the studies extracted in this review report only this single largest-eigenvalue-ratio statistic, with some labeling it ``CI'' and others labeling it ``CN'' more or less interchangeably. Because the quantity that is actually computed is almost always the same — the (square-rooted) ratio between the largest and the smallest eigenvalue — whether a given study calls it a condition index or a condition number is best treated as a naming inconsistency inherited from the literature, rather than as evidence of a different or incorrect diagnostic procedure.

\subsection{Misleading in relation to thresholds of Condition Number}

Of the 100 included studies that report a condition index, condition number, or eigenvalue-based diagnostic, the large majority (61\%) do not state any threshold against which that value is to be interpreted, leaving the reader unable to judge whether the reported figure was itself considered evidence of problematic multicollinearity. Among the minority that do specify a cutoff, the choice of threshold is far from standardized: \cite{review121,review137,review150,review155,review157,review158,ertan2024identifying} compare CN value against 10, 15 studies compare against 30 (\cite{review3}, \cite{review6}, \cite{review8}, \cite{review11}, \cite{review17}, \cite{review23}, \cite{review44}, \cite{review108}, \cite{review116}, \cite{review127}, \cite{review133}, \cite{review151}, \cite{review154}, \cite{review_w6} and \cite{review_w28}) and  six compare against 100 (\cite{review42}, \cite{review47}, \cite{review57}, \cite{review80}, \cite{review83}, \cite{review106}). On the other hand, several authors propose categorical intensity scales rather than a single cutoff point: \cite{review_w34} establishes the values 10 and 30 as boundaries for a three-level scale (none, moderate, and severe), whereas \cite{review42}, \cite{review_w16}, and \cite{review_w20} combine thresholds of 30 (or 15–30) to detect the presence of multicollinearity and 100 to diagnose severe multicollinearity. \cite{review115}, uses a threshold as high as 1000. The arbitrariness of this choice is illustrated most clearly by \cite{review126}, which applies the very same estimator to two different real datasets within the same paper and compares its first application against a threshold of 30 but its second against a threshold of 10, without offering a rationale for the change. A further six studies report no explicit threshold at all but can be inferred, from the qualitative language used to characterize the reported value (e.g., describing it as indicative of ``moderate to strong multicollinearity''), to be implicitly comparing against either 30 or 100 (\cite{review2}, \cite{review5}, \cite{review18}, \cite{review24}, \cite{review25}, \cite{review33}). Taken together, these patterns suggest that the choice of threshold for the condition index/condition number is treated as a largely arbitrary, often unstated convention rather than as a methodologically justified decision.

\subsection{Misleading in relation to essential and non-essential multicollinearity}

To the best of our knowledge, none of the reviewed papers explicitly diagnoses non-essential multicollinearity. Only \cite{review35} and \cite{review39, review121} acknowledge that ill-conditioning
may arise from the intercept, and they do so precisely to justify removing it through standardization, so that non-essential multicollinearity is eliminated from the diagnosis rather than assessed. This omission is particularly relevant because a large part of the reviewed studies compute the diagnostic on a matrix from which the intercept has been removed. Eight studies explicitly compute $\mathbf{X}'\hat{\mathbf{W}}\mathbf{X}$ without the intercept \citep{review8,review33,review25,review37,review42,review108,review81,review123}, and three use $\mathbf{X}'\mathbf{X}$ without the intercept \citep{review3,review6,review126}.
A further eleven studies remove the intercept implicitly, either by centering the covariates or the columns of $\hat{\mathbf{W}}^{1/2}\mathbf{X}$ before the diagnosis, or by fitting the model without it \citep{review74,review19,review155,review137,review35,review66,review39,review121,review114,review83,review40}. Since non-essential multicollinearity is defined precisely by the relationship between the intercept and the explanatory variables, removing the intercept from the analyzed matrix rules out, by construction, the possibility of detecting it. The same holds for \cite{review_w20} and \cite{ertan2024identifying}, who center the covariates and then add a column of ones: the intercept is retained, but it is orthogonal to the regressors by construction. As a result, the diagnosis reported in these 24 studies can be misleading, since a favorable outcome may simply reflect the structural inability of the chosen matrix to reveal non-essential multicollinearity.

\section{Review of methodological proposal for calculating the CN in GLMs}\label{methods}

The previous systematic literature review concluded that, when multicollinearity is assessed in GLMs, the condition number is the most commonly used measure. However, the underlying structure and specific characteristics of GLMs presented in section \ref{glm_structure} pose particular challenges for the detection of multicollinearity, which, as shown in the previous review, are not always taken into account. In this section, we examine the main methodological approaches proposed for the calculation of the Condition Number (CN) in these models.

First, note that the extension of the CN to GLM is based on the application to OLS presented by \cite{Belsley1980} which is based on the singular value decomposition (SVD) of the data matrix $\mathbf{X}$, providing a numerically stable framework for identifying and assessing near-linear dependencies among the explanatory variables. Unlike diagnostics based directly on the eigenvalues of the cross-product matrix $\mathbf{X}'\mathbf{X}$, the SVD operates directly on the data matrix and is computationally more stable. The singular values of $\mathbf{X}$ are related to the eigenvalues of $\mathbf{X}'\mathbf{X}$ through their square roots.

Let $\upsilon_1\geq \upsilon_2\geq\cdots\geq \upsilon_p>0$ denote the singular values of the appropriately scaled design matrix $\mathbf{X}$. The condition index associated with the $i$-th singular value is defined as $CI=\frac{\upsilon_{max}}{\upsilon_{i}}$ for $i=1,...,p$ where $\upsilon_{max}$ is the largest singular value. The condition number is the largest condition index and is therefore given by $CN=\frac{\upsilon_{max}}{\upsilon_{min}}$ where $\upsilon_{min}$ is the smallest singular value. \cite{Belsley1980} concluded that weak dependencies are associated with condition number around 5 or 10, whereas moderate to strong relations are associated with condition number of 30 to 100.

When the diagnostic is computed from the eigenvalues of the cross-product matrix $\mathbf{X}'\mathbf{X}$, denoted by $\lambda_1 \geq \cdots \geq \lambda_p$, the corresponding condition number is
$$\mathrm{CN}_{OLS}=\sqrt{\frac{\lambda_{\max}}{\lambda_{\min}}},$$
because $\upsilon_j=\sqrt{\lambda_j}$. Therefore, the commonly used threshold of 10 for detecting substantial multicollinearity when working with the eigenvalues of $\mathbf{X}'\mathbf{X}$ is the root squared counterpart of the threshold of 100 on the corresponding singular-value scale. Consequently, a large condition number indicates that the design matrix is close to singularity and that the estimated coefficients may be sensitive to small perturbations in the data. Following \cite{Belsley1980}, the variables are first transformed to unit length, i.e., each column of the design matrix \(\mathbf{X}\) is transformed to $\mathbf{X}_{ul}$ so that $\|(x_{ul})_j\| = \sum_{i=1}^{n} ((x_{ul})_{ij})^{2} = 1, \quad j=1,\dots,p$, where \(p\) is the number of predictors. This unit-length scaling ensures that the condition index is not distorted by differences in variable scales, while being equal to 1 in the complete absence of multicollinearity.\\

Different methodologies have been proposed for the extension of the condition number in GLMs. First, \cite{mackinnon1989collinearity}  normalizes the columns of the design matrix $\mathbf{X}$ to unit length ($\mathbf{X}_{ul}$). Upon convergence of the Iteratively Reweighted Least Squares (IRLS) algorithm, they construct the expected information matrix $I(\hat{\boldsymbol{\beta}})_{MP} = \mathbf{X}_{ul}'\hat{\mathbf{W}}\mathbf{X}_{ul}$. Using the Singular Value Decomposition (SVD) of the transformed matrix $\hat{\mathbf{W}}^{1/2}\mathbf{X}_{ul}$, they calculate the condition number as:
\begin{equation}
    CN_{MP} = \sqrt{\frac{\lambda_{\max}(\mathbf{I}(\hat{\boldsymbol{\beta}})_{MP})}{\lambda_{\min}(\mathbf{I}(\hat{\boldsymbol{\beta}})_{MP})}}
\end{equation}
They propose a practical threshold of 10 to declare the presence of damaging multicollinearity in GLMs.

In contrast, \cite{weissfeld1991multicollinearity} proposed to fit the GLM directly on the untransformed matrix $\mathbf{X}$. Once the model converges, they extract the expected information matrix $I(\hat{\boldsymbol{\beta}}) = \mathbf{X}' \hat{\mathbf{W}}\mathbf{X}$ and perform a post-hoc scaling using $S= \{\text{diag}[\mathbf{I}(\hat{\boldsymbol{\beta}})]\}^{-1/2}$, creating the unit length expected information matrix  $I(\hat{\boldsymbol{\beta}})_{WS} = \mathbf{S}\cdot\mathbf{I}(\hat{\boldsymbol{\beta}})\cdot\mathbf{S}$ where all diagonal elements equal 1. They then execute a SVD on $I(\hat{\boldsymbol{\beta}})_{WS}$ to derive the condition number:

\begin{equation}
    CN_{WS} = \sqrt{\frac{\lambda_{\max}(I(\hat{\boldsymbol{\beta}})_{WS})}{\lambda_{\min}(I(\hat{\boldsymbol{\beta}})_{WS})}}
\end{equation}
Note that this transformation is equivalent to transform $\hat{\mathbf{W}}^{1/2}\mathbf{X}$ to unit length, it is to say $I(\hat{\boldsymbol{\beta}})_{WS}=(\hat{\mathbf{W}}^{1/2}\mathbf{X})'_{ul}(\hat{\mathbf{W}}^{1/2}\mathbf{X})_{ul}$. see Appendix B. Based on empirical studies, \cite{weissfeld1991multicollinearity} established condition index cutpoints of 30 for binary logistic regression and 60 for proportional odds models (which fit multiple intercept terms). \cite{marx1990weighted} proposed another condition number applying centering in addition to unit length transformation.

\cite{review133} rigorously analyzed the geometry of the information matrix and support \cite{mackinnon1989collinearity} expected Hessian approach. They concluded that only the condition number based on $\mathbf{I}(\hat{\boldsymbol{\beta}})_{MP}$ is structurally interpretable because it represents the actual inflation of the parameter estimate variances. They demonstrate that if the diagonal elements $w_{ii}$ of the weight matrix are reasonably homogeneous, the multicollinearity in the GLM matches that of the Standard Linear Model (SLM). However, if the weights are highly unequal (e.g., in Poisson or Gamma models with extreme predictions), the condition number can widen dramatically, causing severe ill-conditioning even if the design matrix is perfectly orthogonal.  They proposed distinguishing between ML-collinearity, which occurs when the ratio between the condition number of the weighted information matrix ($CN_{MP}$) and the condition number of the unweighted design matrix ($CN_{OLS}$) exceeds 5 (with $CN_{MP} > 30$), and intrinsic multicollinearity among the explanatory variables, which occurs when $CN_{OLS} > 30$. They also noted that the ratio may be less than 1, indicating the presence of multicollinearity in $\mathbf{X}_{ul}'\mathbf{X}_{ul}$ but not in $\mathbf{X}_{ul}'\hat{\mathbf{W}}\mathbf{X}_{ul}$.

\cite{review137} proposed to diagnose and mitigate multicollinearity introducing a perturbation matrix $\mathbf{V}$ (symmetric and with all diagonals zero) that it is adding to the fisher information matrix $\mathbf{X}'\hat{\mathbf{W}}\mathbf{X}+\mathbf{V}$. While traditional transformations  are linear transformations that preserve the structure underlying the probability of the model, the additive perturbation of \cite{review137} artificially alters the curvature of the log-likelihood function at the optimal point. In addition, given that the trace is equal to the sum of its eigenvalues, the matrix $\mathbf{V}$ necessarily has both positive and negative eigenvalues, which classifies it mathematically as a strictly indefinite matrix. Adding an indefinite matrix to the Fisher information may destroy its positive definiteness, and even when
it does not, the perturbed matrix no longer corresponds to the curvature of the log-likelihood at the maximum likelihood estimate.

\begin{figure}[htbp]
\centering

\begin{tikzpicture}[
    node distance=7mm and 6mm,
    header/.style={
        rectangle,
        draw,
        fill=gray!20,
        rounded corners=2pt,
        align=center,
        font=\bfseries\sffamily\small,
        text width=6cm,
        inner sep=4pt
    },
    box/.style={
        rectangle,
        draw,
        rounded corners=3pt,
        align=center,
        font=\sffamily\footnotesize,
        text width=6cm,
        inner sep=5pt
    },
    mathbox/.style={
        rectangle,
        draw,
        fill=gray!5,
        rounded corners=3pt,
        align=center,
        font=\sffamily\footnotesize,
        text width=6cm,
        inner sep=5pt
    },
    result/.style={
        rectangle,
        draw,
        fill=gray!12,
        rounded corners=3pt,
        align=center,
        font=\sffamily\footnotesize,
        text width=6cm,
        inner sep=5pt
    },
    arrow/.style={
        draw,
        thick,
        -{Stealth[length=2mm]}
    }
]


\node (MP0) [header] {
    MacKinnon \& Puterman (1989)\\
    \textnormal{Design Space Scaling}
};

\node (MP1) [box, below=of MP0] {
    \textbf{1. Scaling \& Model Fitting (IRLS)}\\[1mm]
    Column-normalize $\mathbf{X}$ to unit length:
    $\mathbf{X}_{ul}$\\
    Fit the GLM on $\mathbf{X}_{ul}$ to obtain the Fisher
    weight matrix $\hat{\mathbf{W}}$
};

\node (MP2) [mathbox, below=of MP1] {
    \textbf{2. Expected Information Matrix}\\[1mm]
    $\mathbf{I}(\hat{\boldsymbol{\beta}})_{MP}
    =\mathbf{X}_{ul}'\hat{\mathbf{W}}\mathbf{X}_{ul}$\\
    \scriptsize
    SVD on
    $\hat{\mathbf{W}}^{1/2}\mathbf{X}_{ul}$
};

\node (MP3) [result, below=of MP2] {
    \textbf{3. Diagnostic \& Threshold}\\[1mm]
    $\mathrm{CN}_{MP}=
    \sqrt{
    \lambda_{\max}(\mathbf{I}_{MP})
    /
    \lambda_{\min}(\mathbf{I}_{MP})
    }$\\
    \textbf{Criterion:} damaging multicollinearity if
    $\mathrm{CN}>10$
};


\node (WS0) [header, right=of MP0] {
    Weissfeld \& Sereika (1991)\\
    \textnormal{Post-Hoc Information Scaling}
};

\node (WS1) [box, below=of WS0] {
    \textbf{1. Raw Data Model Fitting (IRLS)}\\[1mm]
    Fit the GLM directly on the raw, uncentered variables
    $\mathbf{X}$\\
    Extract the expected Information Matrix:
    $\mathbf{I}(\hat{\boldsymbol{\beta}})
    =\mathbf{X}'\hat{\mathbf{W}}\mathbf{X}$
};

\node (WS2) [mathbox, below=of WS1] {
    \textbf{2. Post-Hoc Scaling}\\[1mm]
    $\mathbf{S}=
    \operatorname{diag}
    \bigl[\mathbf{I}(\hat{\boldsymbol{\beta}})\bigr]^{-1/2}$\\
    $\mathbf{I}(\hat{\boldsymbol{\beta}})_{WS}
    =\mathbf{S}\mathbf{I}(\hat{\boldsymbol{\beta}})\mathbf{S}$\\
    $(\operatorname{diag}=1)$
};

\node (WS3) [result, below=of WS2] {
    \textbf{3. Diagnostics \& Thresholds}\\[1mm]
    $\mathrm{CN}_{WS}=
    \sqrt{
    \lambda_{\max}(\mathbf{I}_{WS})
    /
    \lambda_{\min}(\mathbf{I}_{WS})
    }$\\
    \textbf{Criterion:} $\mathrm{CN}>30$ (Logistic) /
    $\mathrm{CN}>60$ (Prop. Odds)};


\draw [arrow] (MP0) -- (MP1);
\draw [arrow] (MP1) -- (MP2);
\draw [arrow] (MP2) -- (MP3);

\draw [arrow] (WS0) -- (WS1);
\draw [arrow] (WS1) -- (WS2);
\draw [arrow] (WS2) -- (WS3);


\draw[
    black!50,
    dashed,
    rounded corners=5pt
]
(MP0.north west) rectangle (MP3.south east);

\draw[
    black!50,
    dashed,
    rounded corners=5pt
]
(WS0.north west) rectangle (WS3.south east);

\end{tikzpicture}

\caption{Schematic comparison of the design-space scaling procedure of MacKinnon and Puterman (1989) and the post-hoc information-matrix scaling procedure of Weissfeld and Sereika (1991).}

\label{fig:comparison_mackinnon_weissfeld}

\end{figure}
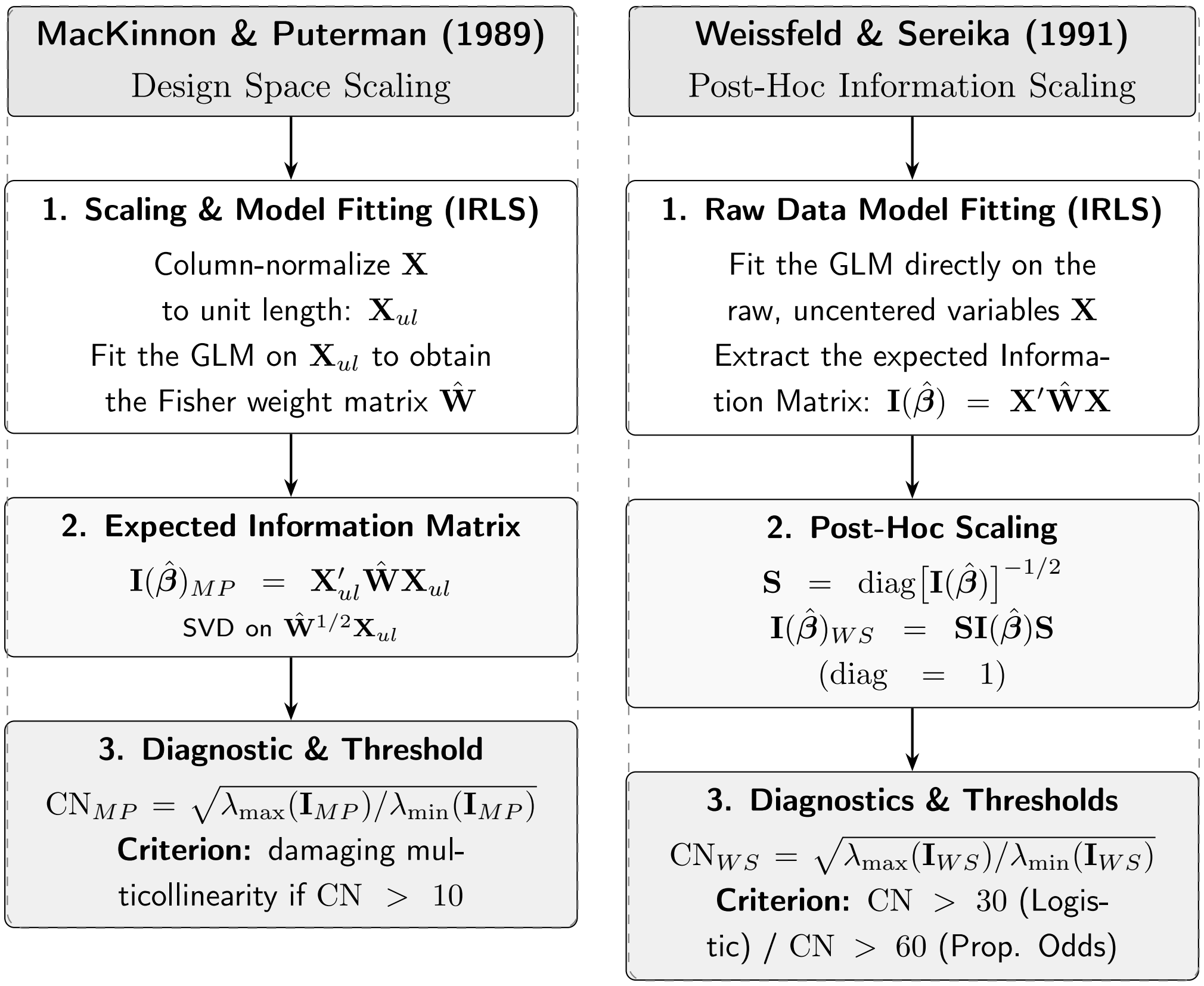

\cite{review42} approaches the problem from a different perspective, suggesting that centering is beneficial when combined with appropriate scaling in GLMs and proposing to remove the intercept from the beginning. The methodology is similar to the one proposed by \cite{mackinnon1989collinearity} but removing the intercept and not only scaling to unit length but also centering. In line with  \cite{Belsley1980} and \cite{weissfeld1991multicollinearity}, we strongly recommend against centering the independent variables, as demeaning the data obscures near-linear dependencies involving the constant intercept. Therefore, the core decision rests on choosing between the uncentered scaling formulations of \cite{weissfeld1991multicollinearity} and  \cite{mackinnon1989collinearity}. Both methodologies are schematically summarized in Figure~\ref{fig:comparison_mackinnon_weissfeld}. In the following subsection, a simple illustrative example is provided to highlight the critical importance of unit-length transformations and to demonstrate why we consider Weissfeld and Sereika's formulation to be the correct approach.

\subsection{An Illustrative Example}
Consider the simple regression model $\mathbf{y} = \beta_0 + \beta_1\mathbf{x} + \mathbf{u}$. In this model, multicollinearity, if present, can only be non-essential, arising from a relationship between the intercept and $\mathbf{x}$, since essential multicollinearity requires at least two explanatory variables. The condition number (CN) should be equal to 1 corresponding to perfect orthogonality between the explanatory variable and the intercept. Suppose that the design matrix is
\[
\mathbf{X} = \begin{bmatrix} \mathbf{1} & \mathbf{x} \end{bmatrix}
= \begin{pmatrix} 1 & 200 \\ 1 & -200 \\ 1 & 20 \\ 1 & -20 \end{pmatrix}.
\]
Since $\mathbf{1}'\mathbf{x} = 200 - 200 + 20 - 20 = 0$, the intercept and the regressor are orthogonal, so no multicollinearity is present. However, computing the condition number from the untransformed data yields:
\[
\mathbf{X}'\mathbf{X} = \begin{pmatrix} 4 & 0 \\ 0 & 80800 \end{pmatrix}
\quad \longrightarrow \quad
CN_{OLS} = \sqrt{\frac{80800}{4}} \approx 142.13,
\]
a value far from its minimum attainable value of 1 and therefore inconsistent with the orthogonality established above. The raw design matrix cannot be used directly; the data must first be rescaled. Any transformation involving centering removes the intercept from the computation and, with it, the ability to detect non-essential multicollinearity. To the best of our knowledge, the only transformation that preserves the intercept is unit-length scaling. Denoting by $\mathbf{X}_{ul}$ the matrix whose columns have been divided by their Euclidean norms ($\sqrt{4} = 2$ and $h = \sqrt{80800}$, respectively),
\[
\mathbf{X}_{ul} = \begin{pmatrix} 1/2 & 200/h \\ 1/2 & -200/h \\ 1/2 & 20/h \\ 1/2 & -20/h \end{pmatrix}
\quad \longrightarrow \quad
\mathbf{X}_{ul}'\mathbf{X}_{ul} = \begin{pmatrix} 1 & 0 \\ 0 & 1 \end{pmatrix}
\quad \longrightarrow \quad
CN_{OLS} = 1.
\]
Unit-length scaling therefore yields $\mathbf{X}_{ul}'\mathbf{X}_{ul} = \mathbf{I}$ under orthogonality, so that the condition number attains its minimum value and correctly identifies the orthogonal case.

Consider now fitting a GLM to the same model. The relevant matrix is no longer $\mathbf{X}'\mathbf{X}$ but $\mathbf{X}'\hat{\mathbf{W}}\mathbf{X}$, with $\hat{\mathbf{W}} = \mathrm{diag}(w_1, w_2, w_3, w_4)$:
\[
\mathbf{X}'\hat{\mathbf{W}}\mathbf{X} = \begin{pmatrix} \sum_{i=1}^4 w_i & a \\ a & b \end{pmatrix},
\qquad
a = \sum_{i=1}^4 w_i x_i = 200(w_1 - w_2) + 20(w_3 - w_4),
\qquad
b = \sum_{i=1}^4 w_i x_i^2 .
\]
The matrix $\mathbf{X}'\hat{\mathbf{W}}\mathbf{X}$ is diagonal if and only if $a = 0$, that is, if and only if the weighted mean of $\mathbf{x}$ is zero, which is equivalent to
\begin{equation}
10\,(w_1 - w_2) = w_4 - w_3 .
\label{eq:orth_condition}
\end{equation}
Equal weights within each pair ($w_1 = w_2$ and $w_3 = w_4$) are sufficient for \eqref{eq:orth_condition}, but not necessary: for instance, $\mathbf{w} = (1.1, 1, 1, 2)$ also gives $a = 20 - 20 = 0$, because the imbalance of the weights in the first pair is compensated by that of the second pair. Unless \eqref{eq:orth_condition} holds, the orthogonality present in $\mathbf{X}'\mathbf{X}$ is lost in $\mathbf{X}'\hat{\mathbf{W}}\mathbf{X}$.

Since using the raw data is already problematic in the linear model, a transformation is also needed in the GLM case, and we again restrict attention to unit-length scaling in order to preserve the intercept. We compare the two proposals of \cite{mackinnon1989collinearity} and \cite{weissfeld1991multicollinearity}, summarized in Figure~2.

\cite{mackinnon1989collinearity} scale $\mathbf{X}$ to unit length before weighting, so that
\[
I(\hat{\boldsymbol{\beta}})_{MP}= \mathbf{X}_{ul}'\hat{\mathbf{W}}\mathbf{X}_{ul}
= \begin{pmatrix} \frac{1}{4}\sum_{i=1}^4 w_i & c \\ c & d \end{pmatrix},
\qquad
c = \frac{a}{2h}, \qquad d = \frac{b}{h^2}.
\]
\cite{weissfeld1991multicollinearity} instead rescale the information matrix itself denoting $I(\hat{\boldsymbol{\beta}})_{WS} = \mathbf{S}\,\mathbf{X}'\hat{\mathbf{W}}\mathbf{X}\,\mathbf{S}$ with $\mathbf{S} = \left\{ \mathrm{diag}\!\left(\mathbf{X}'\hat{\mathbf{W}}\mathbf{X}\right) \right\}^{-1/2}$, which is equivalent (as previously commented) to denote $\mathbf{D} = \mathbf{W}^{1/2}\mathbf{X}$ and transform it to unit length dividing each column of $\mathbf{D}$ by its norm, $\sqrt{h_1}$ and $\sqrt{h_2}$, with $h_1 = \sum_{i=1}^4 w_i$ and $h_2 = b$. Thus, it is verified that $\mathbf{D}_{ul}'\mathbf{D}_{ul}=I(\hat{\boldsymbol{\beta}})_{WS}$:

\[
I(\hat{\boldsymbol{\beta}})_{WS} = \mathbf{S}\,\mathbf{X}'\hat{\mathbf{W}}\mathbf{X}\,\mathbf{S}
= \begin{pmatrix} 1 & e \\ e & 1 \end{pmatrix},
\qquad
e = \frac{a}{\sqrt{h_1 h_2}},
\qquad
\]
which has a unit main diagonal, in line with the convention used under OLS.

\textcolor[rgb]{0.00,0.70,0.00}{\textcolor[rgb]{0.00,0.07,1.00}{}}
Since $c$ and $e$ are both proportional to $a$, the two proposals agree on when orthogonality holds in the GLM metric: $I(\hat{\boldsymbol{\beta}})_{MP}$ and $I(\hat{\boldsymbol{\beta}})_{WS}$ are diagonal if and only if $a = 0$. They differ, however, in how the resulting condition number behaves in that case. Under orthogonality, $I(\hat{\boldsymbol{\beta}})_{WS} = \mathbf{I}$ and $CN_{WS} = 1$ for any weights, whereas
\[
I(\hat{\boldsymbol{\beta}})_{MP} = \begin{pmatrix} o_1 & 0 \\ 0 & o_2 \end{pmatrix},
\qquad
o_1 = \frac{\sum_i w_i}{4}, \qquad o_2 = \frac{\sum_i w_i x_i^2}{\sum_i x_i^2},
\]
and $CN_{MP} = \sqrt{\max(o_1,o_2)/\min(o_1,o_2)}$ equals 1 only if the two weighted averages of the weights coincide. For $\mathbf{w} = (1.1, 1, 1, 2)$, for example, $o_1 = 1.275$ and $o_2 = 85200/80800 \approx 1.054$, so that $CN_{MP} \approx 1.10$ despite exact orthogonality, while $CN_{WS} = 1$.

This result holds in general. For any design with $p$ explanatory variables and an intercept, if $\mathbf{X}'\hat{\mathbf{W}}\mathbf{X}$ is diagonal then $I(\hat{\boldsymbol{\beta}})_{WS}$ is the identity matrix and $CN_{WS} = 1$, whereas $I(\hat{\boldsymbol{\beta}})_{MP}$ is diagonal with elements $o_j = \sum_i w_i x_{ij}^2 / \sum_i x_{ij}^2$, $j = 1, \dots, p$, and $CN_{MP}= 1$ only if all $o_j$ are equal. The diagnostic of \cite{weissfeld1991multicollinearity} therefore attains its minimum value whenever orthogonality holds in the GLM, as the unit-length scaled condition number does in the linear model, whereas the diagnostic of \cite{mackinnon1989collinearity} generally does not. A simulation study, described in the Appendix C, was conducted to assess how frequently orthogonality is lost in practice across the 23 distribution–link combinations considered in Table 2. The results show that orthogonality is preserved only in the four combinations yielding constant weights, whereas it is lost in all replications for the remaining 19 combinations. Although the loss of orthogonality tends to decrease with sample size, it does not disappear, and it generally becomes more pronounced as the number of explanatory variables increases, particularly in small samples.

\section{Condition Number Diagnostics for Multicollinearity in GLMs: Insights from Empirical Case Studies }\label{package}

To assess the practical implications of the methodological issues identified in the literature review, five real-data applications frequently used in previous studies are reanalyzed. For each dataset, we compute and compare the condition number based on the design matrix after the recommended unit-length (UL) transformation (CN$_{OLS}$), as well as the condition number obtained directly from the cross-product matrix ($\mathbf{X}'\hat{\mathbf{W}}\mathbf{X}$) without applying the UL transformation (CN$_W$ (raw)). In addition, we reproduce the condition-number formulations proposed by \cite{mackinnon1989collinearity} (CN$_{MP}$), \cite{weissfeld1991multicollinearity} (CN$_{WS}$), and \cite{review42} (CN$_{OZ}$). This comparative analysis presented in Table \ref{replication} allows us to quantify how the choice of matrix, scaling procedure, and definition of the condition number affects the assessment of multicollinearity. Importantly, the values shown in bold  are those for which the reported results could be successfully replicated. For the remaining results, we were unable to reproduce the values reported in the original studies and instead obtained different values when applying the corresponding computational procedures using the information available for each dataset. In particular, for the Lee dataset, the reported results could be replicated. In the Mine and Body Fat datasets, the condition number reported in the literature corresponds to that calculated from the original, untransformed and weighted design matrix, and these values were also reproduced. For the remaining condition-number formulations reported in these studies, however, we were unable to obtain the published values and instead obtained different values when applying the corresponding computational procedures using the available information.

A subsection is devoted to each dataset. For each application, we describe the original analysis reported in the literature, including the condition-number formulation and computational procedure used by the original authors, and compare their reported results with those obtained under the alternative formulations. Particular attention is given to differences arising from the choice of matrix, the application or omission of the unit-length transformation, and the specific definition of the condition number. The code required to replicate all examples and reproduce the results is publicly available on GitHub (\url{https://github.com/cbgarciaugr/Code_paper_Multicollinearity-diagnosis-in-generalized-linear-models}).

\begin{table}
\caption{Summary of replication results for different datasets}
\label{replication}
\small{
\begin{tabular}{cccccccc}
  \hline
Application  & Distribution (Link)&     Intercept  &  CN$_{OLS}$  &   CN$_W$ (raw)   & CN$_{MP}$ &   CN$_{WS}$   & CN$_{OZ}$\\
\hline
Lee Cancer& Binomial(Logit)& Yes& 190.7765& 10808.1330& \textbf{329.9542} &\textbf{315.1366} & 16.6975\\
Lee Cancer& Binomial(Logit)& No&  11.7696 & 1347.1513 & 33.0269&  32.3722 & 15.3222\\
\hline
Nitrogen& Gamma(inverse)&Yes & 67.7053  &8790.6504 & 75.9081&  77.9772&  76.1047\\
Nitrogen& Gamma(inverse)&No & 14.6751 &  119.2248 & 13.7742&  14.3986  &    Inf\\
\hline
Mine &Poisson (log)& Yes & 19.8950  &\textbf{2093.3158}  &26.4538 & 28.5822&   2.0946\\
Mine &Poisson (log)&No  & 6.4023   & 31.3666 &  6.9798  & 7.4399  & 1.5372\\
\hline
Bodyfat& Gamma(log)&Yes& 352.0413  &\textbf{3410.6317} &352.0413 &352.0413  &31.3229\\
Bodyfat& Gamma(log)&No &333.7805&  3402.5909 &333.7805 &333.7805 & 31.3229\\
\hline
Plastic& Poisson (log)&Yes& 595.7429& 73561.6243& 899.0447& 876.8393& 254.8772\\
Plastic& Poisson (log)&No  &342.7521 &17495.3548 &325.5947 &323.4382 &257.4191\\
  \hline
\end{tabular}}
\end{table}

\subsection{Lee cancer}
The Lee cancer remission dataset, originally compiled by \cite{lee1974computer}, is traditionally applied to examine multicollinearity diagnostics in binomial logistic regression. The dataset consists of clinical observations from 27 patients, of whom 9 experienced complete cancer remission ($Y = 1$) and 18 experienced incomplete remission ($Y = 0$). To model the probability of complete remission, the dataset records six quantitative covariates representing patient characteristics: cell index (CELL), smear index (SMEAR), infiltration index (INFIL), labeling index (LI), blast index (BLAST), and body temperature (TEMP). This dataset is applied by \cite{review35, review42, review133, review137}. \cite{review133} is the first in the selection of reviewed paper that analyzed this data set and use only three explanatory variables LI, TEMP, and CELL that exhibits severe multicollinearity, presenting a classic real-world scenario of extreme variance inflation and numerical instability. The results obtained in Table \ref{replication} coincide exactly with the one reported by \cite{review133}.

\cite{review35} uses the variables cell index (CELL), smear index (SMEAR), infiltration index (INFIL), blast index (BLAST) and body temperature (TEMP). They standardized the variables and explicitly stated ``standardization removes any non-essential ill-conditioning resulting from the intercept''. They obtained a value of 295.7026 for the condition number as directly the ratio between the eigenvalues, and applying the square root the value of 17.196. \cite{review42} also used this data set to obtain the condition number directly from $\mathbf{X}'\hat{\mathbf{W}}\mathbf{X}$ without considering the intercept. Without transformation authors provided the following eigen values: 15.6562, 0.4601, 0.1557, 0.0145, 0.0001, from which they reported a condition number equal to 1.3032 $\times 10^5$. This values is not exactly obtained from that eigenvalues and we guess that is directly the ratio and not the square root. The same occurs when center and scale transformation is applied. They report the following eigenvalues 4.4133, 0.3657; 0.2009, 0.0200 and 0.0002, obtaining a condition number equal to 2.6099 $\times 10^4$. Again, this values is not exactly obtained from that eigenvalues and we guess that is directly the ratio and not the square root. \cite{review137} uses the full data set with all the explanatory variables obtaining a condition number equal to 418.13.

\cite{ertan2024identifying} use the same five variables as \cite{review35} (CELL, SMEAR, INFIL, BLAST and TEMP). They standardize them so that $\mathbf{X}'\mathbf{X}$ is a correlation matrix and report a condition number of 201.33. \cite{review_w16} report exactly the same value, 201.33, for the same dataset after standardizing the predictors so that $\mathbf{X}'\mathbf{X}$ is a correlation matrix. However, they do not state the matrix from which the eigenvalues are obtained, they display six predictors (including LI) instead of five, and they cite \cite{ertan2024identifying} as a source of the data, so the figure seems to be taken from that work rather than recomputed.

\subsection{Nitrogen data set}

The second dataset considered in the replication analysis is the air pollution dataset originally reported by \cite{chatterjee1988sensitivity}. The data comprise measurements of weather conditions and nitrogen dioxide concentrations (y), expressed in parts per hundred million (p.p.h.m.), recorded over September 1984 at a monitoring station in the San Francisco Bay Area. Four explanatory variables are considered: mean wind speed ($x_1$), measured in miles per hour (m.p.h.); maximum temperature ($x_2$), measured in degrees Fahrenheit; insolation ($x_3$), measured in langleys per day; and stability factor ($x_4$), measured in degrees Fahrenheit. Although \cite{chatterjee1988sensitivity} describes the dataset as containing 26 observations, inspection of the original data reveals that the observation numbering jumps from observation 8 to observation 10, with no observation 9 reported. Consequently, the dataset actually contains 25 observations rather than 26. To the best of our knowledge, this discrepancy has not been previously identified or corrected in the subsequent literature using this dataset. In the present study, we use the 25 observations actually reported in the original source and explicitly account for this discrepancy in the replication analysis.

Among the reviewed articles, the first to use this dataset is \cite{review69}, who fit a gamma model with reciprocal link and intercept and state that, ``since the units of the explanatory variables are different, we standardized the explanatory variables''. Then the constant term is added to the $X$ matrix. They compute the condition number on $\mathbf{X}'\hat{\mathbf{W}}\mathbf{X}$ as the ratio between the largest and the
smallest eigenvalue (without taking the square root), obtaining 213.8097, which they take as evidence of multicollinearity. This value is subsequently reproduced by most later studies (\cite{review2}, \cite{review16}, \cite{review34}, \cite{review94}, \cite{review118}) without being recalculated. However, in two cases (\cite{review34} and \cite{review94}) the fitted model itself uses a logarithmic link rather than the reciprocal link that originally produced the value, so the condition number they report almost certainly does not correspond to their own fitted weight matrix. \cite{review97} performs an analogous calculation but with the IRLS weights of their own log-gamma model, obtaining a substantially larger condition number of 49649929. \cite{review47} applies the same dataset under an Inverse Gaussian distribution (not Gamma) with a log link, carrying out a similar calculation and obtaining 10263.65. \cite{review42} and \cite{review66} are the only two studies that document a full transformation of the data for the multicollinearity diagnostics. \cite{review66} weights the design matrix by $\hat{\mathbf{W}}^{1/2}$, centers it, and rescales each column to unit length, obtaining a condition number of $2.49\times10^{8}$; the authors also compute the VIF via a Moore–Penrose pseudoinverse, since the resulting matrix is nearly singular, and note themselves that this is why their VIF values come out artificially low (0.19–5.49) despite the astronomically large condition number. \cite{review42}, using the same weighting/centering/unit-length-scaling procedure, reports both an untransformed scenario (CN $=1.4820\times10^{4}$) and a transformed one (CN $=1.5508\times10^{16}$, driven by a single eigenvalue that is numerically zero), making explicit how strongly the diagnostic depends on this choice. In summary, only \cite{review42} and \cite{review66} center and scale the weighted matrix for the collinearity diagnostic; the remaining studies work directly on the original variables weighted by $\hat{\mathbf{W}}$, and the ``standard'' figure of 213.8097 that dominates this literature is, for the most part, an inherited citation rather than an independent computation.

We were unable to reproduce the values reported in \cite{review69}, and subsequently in \cite{review2} ($\hat\phi^2 = 0.07572852$, CN $= 213.8097$); refitting the model under several standardizations consistently yields $\hat\varphi^2\approx0.1126$, since this estimate is invariant to any rescaling of the design matrix, suggesting an unreported detail in their estimation procedure rather than a data or scaling issue.

The contrast between the two rows of Table~1 for this dataset clarifies the origin of the
condition number of order $10^{16}$ reported by \cite{review42}. When the model includes an
intercept, the formulation of \cite{review42} yields $CN_{OZ}=76.10$, very close to
$CN_{WS}=77.98$. When the model is fitted without an intercept, however, $CN_{OZ}$ is
infinite. This is not a numerical accident but an exact algebraic consequence of the
canonical (inverse) link of the gamma model. Since $\eta_i = \mathbf{x}_i'\boldsymbol\beta = 1/\mu_i$ and $w_{ii} = \mu_i^2/\phi$,
it follows that
\begin{equation*}
\sum_{j=1}^{p} \hat\beta_j \sqrt{\hat w_{ii}}\, x_{ij} = \sqrt{\hat w_{ii}}\,\hat\eta_i
= \frac{1}{\sqrt{\phi}}, \qquad i=1,\dots,n,
\end{equation*}
that is, a linear combination of the weighted regressors is exactly constant. Centering the
weighted columns maps this constant vector to zero, so the centered weighted matrix is
exactly rank deficient. The value $1.55\times 10^{16}$ reported by
\cite{review42} is therefore an artifact of combining centering of the weighted regressors
with a model fitted without an intercept, rather than a measure of the severity of the
multicollinearity in the data. 

\subsection{Mines}
The mine fracture dataset — 44 observations on roof fractures recorded at coal mines in the Appalachian region of West Virginia, with four continuous predictors describing inner burden thickness, extraction percentage of the previously mined lower seam, lower seam height, and time since the mine was opened — is the real-data application shared by five of the reviewed articles, each fitting a different count-data GLM to it. \cite{review55} and \cite{review118}, both by the same authors, fit a standard Poisson model with a log link and propose, respectively, a restricted ridge estimator and a restricted Liu estimator; \cite{review8} fits a Conway–Maxwell-Poisson model to accommodate the dispersion observed in the data and proposes a ridge estimator (COMPRR); \cite{review115} fits a Bell regression model and proposes a Kibria–Lukman-type estimator; and \cite{review117} fits a discrete Weibull regression model and proposes a ridge estimator (DWRE). \cite{review8} state that they analyze a different distribution but do not provide any diagnostic measures for the mine data (which should differ across distributions), instead referencing a previous paper to claim that the dataset presents multicollinearity. \cite{review117} likewise provide no diagnostic measures and cite \cite{review8} to justify the existence of multicollinearity, despite using a different distribution. In addition, from the code provided in the appendix of \cite{review117}, it appears that the diagnostic is computed on the unweighted $\mathbf{X}'\mathbf{X}$; moreover, although the code includes a standardization step, it is never actually applied. \cite{review55} and \cite{review118}, applying the Poisson distribution with a log link, obtain a condition number equal to 2093.32 that is the one obtained when the CN is calculated from the untransformed weighted matrix $\mathbf{X}'\hat{\mathbf{W}}\mathbf{X}$, see Table \ref{replication}. \cite{review115}, applying the Bell distribution, obtains a CN equal to 2139.89.

When the data are unit-length scaled (without centering) and the intercept is included (Poisson family, log link), the resulting condition number  (obtained as the square root of the ratio between the maximum and the minimum eigenvalue of the matrix $\mathbf{X}'\hat{\mathbf{W}}\mathbf{X}$) is equal to 28.58. Although the conclusion regarding the presence of multicollinearity remains affirmative in this case, it is important to note that the value is much lower, and that in other cases the conclusion could change.

\subsection{Body fat}

The bodyfat dataset from the \texttt{TH.data} package \citep{hothorn2019package}, originally presented by \cite{garcia2005improved}, contains information on 71 healthy German women and includes measurements related to body composition and anthropometry. The main response variable is DEXfat, which represents body fat measured using dual-energy X-ray absorptiometry (DXA). The dataset also includes age, waist circumference, hip circumference, elbow breadth, knee breadth, and several anthropometric variables (anthro3a, anthro3b, anthro3c, and anthro4) derived from combinations of body measurements. The main purpose of the dataset is to investigate whether body fat can be predicted from simpler anthropometric measurements.

This dataset has been analyzed by \cite{review23, review44, review51, review76, review127}, all of whom report having calculated the condition number as the square root of the ratio between the maximum and minimum eigenvalues of $\mathbf{X}'\hat{\mathbf{W}}\mathbf{X}$, under a Gamma distribution assumption. \cite{review23, review44, review127} report an identical condition number (CN = 3394.30); however, none of them explicitly states which link function was applied, so we were unable to reproduce that value. We reproduce the value reported by \cite{review51} (CN = 3410.63) which is obtained when a log link function is applied and the data are not transformed, see Table \ref{replication}. \cite{review76} explicitly states that a log link function is also applied, yet obtains a different condition number (CN = 4026.23).

When the data are unit-length scaled (without centering) and the intercept is included, the resulting condition number, obtained as the ratio between the maximum and the minimum eigenvalue of the matrix $\mathbf{X}'\hat{\mathbf{W}}\mathbf{X}$, is equal to 352.04. Although the conclusion regarding the presence of multicollinearity remains affirmative in this case, it is important to note that, in other cases, the choice of calculation method can affect the resulting conclusions.

\subsection{Plastic}

The plastic plywood process dataset, utilized as a real-world case study by \cite{marcondesfilho2016principal} in the context of statistical process control, originates from a medium-sized timber industry that manufactures laminated plastic plywood. The dataset consists of 100 reference observations designed to evaluate how specific manufacturing parameters influence the final product's quality. The response variable ($y$) represents the number of defects or imperfections found per laminated plastic plywood area, which is modeled as count data. The process is evaluated through four quantitative input variables (covariates): volumetric shrinkage ($x_1$, in \%), assembly time ($x_2$, in minutes), wood density ($x_3$, in $g/cm^3$), and drying temperature ($x_4$, in $^{\circ}$C). This dataset is particularly renowned in collinearity literature because the input variables exhibit strong, statistically significant pairwise correlations—specifically between shrinkage and assembly time ($x_1$ and $x_2$), as well as wood density and drying temperature ($x_3$ and $x_4$). This severe multicollinearity causes standard Poisson regression models to yield highly unstable and statistically insignificant coefficient estimates.

This dataset has been analyzed by \cite{review8,review19,review27,review36,review74,review81}. \cite{review19} and \cite{review74} applied the Bell regression without intercept with centered and standardized data obtaining a CN equal to 74.5281. Note that the transformation is performed on the original data, since \cite{review74} explicitly said ``The design matrix is centred and standardized so that $\mathbf{X}'\mathbf{X}$ is in the correlation form''.

\cite{review8}  applied the COM Poisson regression model calculating the condition number as the square root of the ratio of eigen values of matrix $\mathbf{X}'\hat{\mathbf{W}}\mathbf{X}$ excluding intercept with a value of 8634.73. No transformation is explicitly specified. \cite{review81} reported the same condition number although they applied not only the COM Poisson but also the Poisson and the negative binomial distribution.

\cite{review36} only used the explanatory variables: assembly time ($x_2$), wood density ($x_3$), and drying temperature ($x_4$) with the COM Poisson model and log link function. The condition number reported is equal to 13120.14 and no more details about transformation or calculation are included.

\cite{review27} reported a condition number equal to 2.9558$\times 10^5$ as the ratio between the maximum and the minimum eigenvalue of matrix $\mathbf{X}'\hat{\mathbf{W}}\mathbf{X}$. No distribution, function link or transformation is explicitly reported.

\section{Conclusions} \label{conclusions}
This study pursued three complementary objectives: first, to systematically review which methodologies are used in practice to diagnose multicollinearity in real-data applications in GLMs, and how they are implemented; second, to review the methodological proposals for computing the condition number in GLMs and identify the most appropriate one; and third, to reassess selected applications from the literature using a diagnostic framework consistent with the underlying GLM structure. Taken together, the findings indicate that, despite the substantial and rapidly growing body of research inspired by \cite{mackinnon1989collinearity}, \cite{weissfeld1991multicollinearity} and \cite{segerstedt1992ordinary}, the empirical justification for applying ridge-type estimators is not always supported by a rigorous assessment of multicollinearity.

The systematic review of 120 studies reveals that the assessment of multicollinearity does not follow a common criterion. Eight studies (6.67\%) applied a ridge-type estimator without reporting any multicollinearity diagnostic, while a further eleven (9.17\%) relied exclusively on pairwise correlations and/or variance inflation factors, which do not provide a global assessment of the conditioning of the model and cannot distinguish essential from non-essential multicollinearity. Among the remaining studies, the condition number is by far the most widely used diagnostic, but it is computed in very different ways: some studies transform the data before the diagnosis and others do not; among those that do, some transform the design matrix and others the weighted information matrix; some compute the diagnostic on the unweighted cross-product matrix $\mathbf{X}'\mathbf{X}$ rather than on the GLM information matrix $\mathbf{X}'\hat{\mathbf{W}}\mathbf{X}$; and thresholds are frequently unstated or inconsistent with the definition of the condition number being used, in particular when the ratio of eigenvalues is reported without taking its square root. The terms condition index and condition number are also used interchangeably. Furthermore, none of the reviewed studies explicitly diagnoses non-essential
multicollinearity. In 24 of them the intercept was either removed from the matrix used for diagnosis, by excluding it, by centering the covariates or the weighted design matrix, or by fitting the model without it, or retained but made orthogonal to the regressors by construction, so that non-essential multicollinearity could not be detected. The only studies that refer to ill-conditioning arising from the intercept \citep{review35,review39,review121} do so precisely to justify removing it through standardization, so that this source of ill-conditioning is eliminated from the diagnosis rather than assessed.

Given that the condition number is the dominant diagnostic, we reviewed the main proposals for its extension to GLMs, namely those of \cite{mackinnon1989collinearity} and \cite{weissfeld1991multicollinearity}. Our analysis leads us to recommend the procedure of \cite{weissfeld1991multicollinearity}, which computes the condition number from the information matrix $\mathbf{X}'\hat{\mathbf{W}}\mathbf{X}$ rescaled to unit length, so that all its diagonal elements equal one. This recommendation rests on several arguments. First, the diagnostic is computed on the matrix that actually determines the variance of the maximum likelihood estimator in a GLM, rather than on the unweighted design matrix. Second, by analogy with the linear model, the resulting condition number attains its minimum value of one whenever $\mathbf{X}'\hat{\mathbf{W}}\mathbf{X}$ is diagonal, that is, under orthogonality in the GLM metric; by contrast, the proposal of \cite{mackinnon1989collinearity}, which scales the design matrix before weighting, does not in general attain this minimum even when orthogonality holds. Third, unit-length scaling without centering preserves the intercept, so the diagnostic remains able to detect non-essential multicollinearity. This is the main limitation of the proposal of \cite{review42}, which weights, centers and scales the regressors: centering removes the intercept from the analyzed matrix and therefore rules out, by construction, the detection of non-essential multicollinearity. Moreover, centering the weighted regressors can push the smallest eigenvalue to machine precision, as illustrated by the nitrogen dioxide data, where the formulation of \cite{review42} yields a condition number of the order of $10^{16}$ (for the model without intercept), whereas the procedure of \cite{weissfeld1991multicollinearity} produces a moderate but well-defined value.

The empirical reassessment of five real-data applications confirms that these issues are not merely matters of incomplete reporting. The choice of matrix, scaling procedure, definition of the diagnostic and treatment of the intercept can alter the numerical assessment of multicollinearity by several orders of magnitude and, consequently, the conclusion about whether ridge estimation is warranted. Published condition numbers can differ substantially from those obtained under the recommended framework, and in some cases apparently extreme values reflect particular transformations or matrix formulations rather than a meaningful measure of the severity of multicollinearity.

Based on these findings, we recommend that empirical applications of ridge and ridge-type GLM estimators adopt a clearly defined and reproducible diagnostic procedure: the condition number should be computed, following \cite{weissfeld1991multicollinearity}, from the unit-length scaled information matrix $\mathbf{X}'\hat{\mathbf{W}}\mathbf{X}$ with the intercept retained; the definition of the condition number (with or without the square root) and the threshold used for its interpretation should be clearly stated and mutually consistent; and any transformation applied to the data should be explicitly reported. These practices would improve both the reproducibility of published results and the empirical justification for the subsequent use of ridge estimation.

Several avenues for future research remain. The citation-based search strategy used in this study was anchored to three index studies and conducted using the Web of Science Core Collection. Future reviews could extend the search to additional bibliographic databases and complementary search strategies to determine whether the methodological patterns identified here persist in a broader body of literature. The empirical reassessment could likewise be extended to a larger collection of real-data applications and to other ridge-type and shrinkage estimators.

\section*{Appendix}

\subsection*{Appendix A}\label{appendixa}

\begin{table}[H]
\centering
\caption{Main distributions, parametrizations, link functions and IRLS
weights used in GLMs and related regression models.}
\label{tab:distributions_links}
\begin{tabular}{p{2.8cm} p{8.0cm} p{7.0cm} p{8.0cm}}
\toprule
\textbf{Distribution} &
\textbf{Full parametrization} &
\textbf{Link function} &
\textbf{IRLS weight $w_{ii}$} \\
\midrule


\multirow{3}{*}{Exponential}
&
\multirow{3}{*}{
$Y_i\sim\operatorname{Exp}(\lambda_i)$,
$\lambda_i>0$
}
&
Inverse:
$\displaystyle g(\mu)=\frac{1}{\mu}$
&
$\displaystyle
\frac{\mu_i^2}{\phi}
$
\\

& &
Log:
$\displaystyle g(\mu)=\log(\mu)$
&
$\displaystyle
\frac{1}{\phi}
$
\\

& &
Identity:
$\displaystyle g(\mu)=\mu$
&
$\displaystyle
\frac{1}{\phi\mu_i^2}
$
\\

\midrule


\multirow{3}{*}{Gamma}
&
\multirow{3}{*}{
$Y_i\sim\operatorname{Gamma}(a_i,b_i)$,

$\displaystyle a_i=\frac{1}{\phi}$,

$\displaystyle b_i=\frac{1}{\phi\mu_i}$
}
&
Inverse:
$\displaystyle g(\mu)=\frac{1}{\mu}$
&
$\displaystyle
\frac{\mu_i^2}{\phi}
$
\\

& &
Log:
$\displaystyle g(\mu)=\log(\mu)$
&
$\displaystyle
\frac{1}{\phi}
$
\\

& &
Identity:
$\displaystyle g(\mu)=\mu$
&
$\displaystyle
\frac{1}{\phi\mu_i^2}
$
\\

\midrule


\multirow{3}{*}{Inverse Gaussian}
&
\multirow{3}{*}{
$Y_i\sim IG(\mu_i,\lambda)$,

$\mu_i>0,\lambda>0$
}
&
Inverse-square:
$\displaystyle g(\mu)=\frac{1}{\mu^2}$
&
$\displaystyle
\frac{\mu_i^3}{4\phi}
$
\\

& &
Inverse:
$\displaystyle g(\mu)=\frac{1}{\mu}$
&
$\displaystyle
\frac{\mu_i}{\phi}
$
\\

& &
Log:
$\displaystyle g(\mu)=\log(\mu)$
&
$\displaystyle
\frac{1}{\phi\mu_i}
$
\\

\midrule


\multirow{3}{*}{Poisson}
&
\multirow{3}{*}{
$Y_i\sim\operatorname{Poisson}(\mu_i)$,

$\mu_i>0$
}
&
Log:
$\displaystyle g(\mu)=\log(\mu)$
&
$\displaystyle
\mu_i
$
\\

& &
Identity:
$\displaystyle g(\mu)=\mu$
&
$\displaystyle
\frac{1}{\mu_i}
$
\\

& &
Square-root:
$\displaystyle g(\mu)=\sqrt{\mu}$
&
$\displaystyle
4
$
\\

\midrule


\multirow{4}{*}{Binomial}
&
\multirow{4}{*}{
$Y_i\sim\operatorname{Binomial}(n_i,p_i)$,

$p_i=\mu_i$
}
&
Logit:
$\displaystyle
g(\mu)=
\log\left(\frac{\mu}{1-\mu}\right)$
&
$\displaystyle
n_i\mu_i(1-\mu_i)
$
\\

& &
Probit:
$\displaystyle g(\mu)=\Phi^{-1}(\mu)$
&
$\displaystyle
n_i
\frac{\varphi(\eta_i)^2}
{\mu_i(1-\mu_i)}
$
\\

& &
Comp.log-log:
$\displaystyle
g(\mu)=\log[-\log(1-\mu)]$
&
$\displaystyle
n_i
\frac{
e^{2\eta_i-2e^{\eta_i}}
}{
\mu_i(1-\mu_i)
}
$
\\

& &
Log:
$\displaystyle g(\mu)=\log(\mu)$
&
$\displaystyle
n_i\frac{\mu_i}{1-\mu_i}
$
\\

\midrule


\multirow{3}{*}{Negative Binomial}
&
\multirow{3}{*}{
$Y_i\sim NB(r,p_i)$,

$r>0$,

$\displaystyle
p_i=\frac{r}{r+\mu_i}$,

$\displaystyle
\alpha=\frac{1}{r}$
}
&
Log:
$\displaystyle g(\mu)=\log(\mu)$
&
$\displaystyle
\frac{\mu_i}
{1+\alpha\mu_i}
$
\\

& &
Identity:
$\displaystyle g(\mu)=\mu$
&
$\displaystyle
\frac{1}
{\mu_i+\alpha\mu_i^2}
$
\\

& &
Canonical:
$\displaystyle
g(\mu)=
\log\left(
\frac{\mu}{1+\alpha\mu}
\right)$
&
$\displaystyle
\mu_i(1+\alpha\mu_i)
$
\\

\midrule


\multirow{3}{*}{Bell}
&
\multirow{3}{*}{
$Y_i\sim\operatorname{Bell}(\theta_i)$,

$\displaystyle
\mu_i=\theta_i e^{\theta_i}$,

$\displaystyle
\theta_i=W(\mu_i)$
}
&
Log:
$\displaystyle g(\mu)=\log(\mu)$
&
$\displaystyle
\frac{\mu_i}
{1+W(\mu_i)}
$
\\

& &
Identity:
$\displaystyle g(\mu)=\mu$
&
$\displaystyle
\frac{1}
{\mu_i[1+W(\mu_i)]}
$
\\

& &
Square-root:
$\displaystyle g(\mu)=\sqrt{\mu}$
&
$\displaystyle
\frac{4}
{1+W(\mu_i)}
$
\\
\bottomrule
\end{tabular}
\end{table}

\subsection*{Appendix B}\label{appendixb}
To illustrate the equivalence between both developments, we start from:
\[
\underset{2\times n}{\mathbf{X}^{t}}\,\underset{n\times n}{\mathbf{W}}
=\begin{pmatrix}
1 & 1 & \cdots & 1\\
x_1 & x_2 & \cdots & x_n
\end{pmatrix}
\begin{pmatrix}
w_1 & 0 & \cdots & 0\\
0 & w_2 & \cdots & 0\\
\vdots & \vdots & \ddots & \vdots\\
0 & 0 & \cdots & w_n
\end{pmatrix}
=\begin{pmatrix}
w_1 & w_2 & \cdots & w_n\\
x_1 w_1 & x_2 w_2 & \cdots & x_n w_n
\end{pmatrix},
\]
and, multiplying by $\mathbf{X}$,
\[
\underset{2\times n}{\mathbf{X}^{t}}\,\underset{n\times n}{\mathbf{W}}\,\underset{n\times 2}{\mathbf{X}}
=\begin{pmatrix}
w_1 & w_2 & \cdots & w_n\\
x_1 w_1 & x_2 w_2 & \cdots & x_n w_n
\end{pmatrix}
\begin{pmatrix}
1 & x_1\\
1 & x_2\\
\vdots & \vdots\\
1 & x_n
\end{pmatrix}
=\begin{pmatrix}
\sum_{i=1}^{n} w_i & \sum_{i=1}^{n} x_i w_i\\[4pt]
\sum_{i=1}^{n} x_i w_i & \sum_{i=1}^{n} x_i^{2} w_i
\end{pmatrix}.
\]

The diagonal of this matrix is
\[
\operatorname{diag}\!\left(\mathbf{X}^{t}\mathbf{W}\mathbf{X}\right)
=\begin{pmatrix}
\sum_{i=1}^{n} w_i & 0\\
0 & \sum_{i=1}^{n} x_i^{2} w_i
\end{pmatrix},
\]
and the scaling matrix is defined as
\[
\underset{2\times 2}{\mathbf{S}}
=\left[\operatorname{diag}\!\left(\mathbf{X}^{t}\mathbf{W}\mathbf{X}\right)\right]^{-1/2}
=\left(\left[\operatorname{diag}\!\left(\mathbf{X}^{t}\mathbf{W}\mathbf{X}\right)\right]^{1/2}\right)^{-1}
=\begin{pmatrix}
\dfrac{1}{\sqrt{\sum_i w_i}} & 0\\[10pt]
0 & \dfrac{1}{\sqrt{\sum_i x_i^{2} w_i}}
\end{pmatrix}.
\]

Then,
\[
\underset{2\times 2}{\mathbf{S}}\,\underset{2\times n}{\mathbf{X}^{t}}
=\begin{pmatrix}
\dfrac{1}{\sqrt{\sum_i w_i}} & 0\\[10pt]
0 & \dfrac{1}{\sqrt{\sum_i x_i^{2} w_i}}
\end{pmatrix}
\begin{pmatrix}
1 & 1 & \cdots & 1\\
x_1 & x_2 & \cdots & x_n
\end{pmatrix}
=\begin{pmatrix}
\dfrac{1}{\sqrt{\sum_i w_i}} & \dfrac{1}{\sqrt{\sum_i w_i}} & \cdots & \dfrac{1}{\sqrt{\sum_i w_i}}\\[10pt]
\dfrac{x_1}{\sqrt{\sum_i x_i^{2} w_i}} & \dfrac{x_2}{\sqrt{\sum_i x_i^{2} w_i}} & \cdots & \dfrac{x_n}{\sqrt{\sum_i x_i^{2} w_i}}
\end{pmatrix}.
\]

The scaled information matrix is
\[
I(\hat{\boldsymbol{\beta}})_{WS}
=\mathbf{S}\,\mathbf{X}^{t}\,\mathbf{W}\,\mathbf{X}\,\mathbf{S}
=(\mathbf{S}\mathbf{X}^{t})\,\mathbf{W}\,(\mathbf{S}\mathbf{X}^{t})^{t}.
\]
Expanding first $(\mathbf{S}\mathbf{X}^{t})\mathbf{W}$:
\[
(\mathbf{S}\mathbf{X}^{t})\mathbf{W}
=\begin{pmatrix}
\dfrac{w_1}{\sqrt{\sum_i w_i}} & \dfrac{w_2}{\sqrt{\sum_i w_i}} & \cdots & \dfrac{w_n}{\sqrt{\sum_i w_i}}\\[10pt]
\dfrac{x_1 w_1}{\sqrt{\sum_i x_i^{2} w_i}} & \dfrac{x_2 w_2}{\sqrt{\sum_i x_i^{2} w_i}} & \cdots & \dfrac{x_n w_n}{\sqrt{\sum_i x_i^{2} w_i}}
\end{pmatrix},
\]
and multiplying by $\mathbf{X}\mathbf{S}=(\mathbf{S}\mathbf{X}^{t})^{t}$ we obtain
\begin{equation}\label{eq:info-escalada}
I(\hat{\boldsymbol{\beta}})_{WS}
=\begin{pmatrix}
1 & \dfrac{\sum_i x_i w_i}{\sqrt{\sum_i w_i}\,\sqrt{\sum_i x_i^{2} w_i}}\\[14pt]
\dfrac{\sum_i x_i w_i}{\sqrt{\sum_i w_i}\,\sqrt{\sum_i x_i^{2} w_i}} & 1
\end{pmatrix}.
\end{equation}
The eigenvalue and eigenvector decomposition is then applied to this matrix.

In relation to the formulation via the matrix $\mathbf{D}$ and its comparison with \cite{weissfeld1991multicollinearity}, let $\mathbf{D}=\mathbf{W}^{1/2}\mathbf{X}$, so that
\[
\mathbf{D}^{t}\mathbf{D}=\mathbf{X}^{t}\mathbf{W}^{1/2}\mathbf{W}^{1/2}\mathbf{X}=\mathbf{X}^{t}\mathbf{W}\mathbf{X}=I(\boldsymbol\beta).
\]
Explicitly,
\[
\underset{n\times 2}{\mathbf{D}}
=\begin{pmatrix}
\sqrt{w_1} & 0 & \cdots & 0\\
0 & \sqrt{w_2} & \cdots & 0\\
\vdots & \vdots & \ddots & \vdots\\
0 & 0 & \cdots & \sqrt{w_n}
\end{pmatrix}
\begin{pmatrix}
1 & x_1\\
1 & x_2\\
\vdots & \vdots\\
1 & x_n
\end{pmatrix}
=\begin{pmatrix}
\sqrt{w_1} & \sqrt{w_1}\,x_1\\
\sqrt{w_2} & \sqrt{w_2}\,x_2\\
\vdots & \vdots\\
\sqrt{w_n} & \sqrt{w_n}\,x_n
\end{pmatrix}.
\]

Scaling the columns of $\mathbf{D}$ to unit length,
\[
\underset{n\times 2}{\mathbf{D}_{ul}}
=\begin{pmatrix}
\dfrac{\sqrt{w_1}}{\sqrt{\sum_i w_i}} & \dfrac{\sqrt{w_1}\,x_1}{\sqrt{\sum_i x_i^{2} w_i}}\\[10pt]
\dfrac{\sqrt{w_2}}{\sqrt{\sum_i w_i}} & \dfrac{\sqrt{w_2}\,x_2}{\sqrt{\sum_i x_i^{2} w_i}}\\
\vdots & \vdots\\
\dfrac{\sqrt{w_n}}{\sqrt{\sum_i w_i}} & \dfrac{\sqrt{w_n}\,x_n}{\sqrt{\sum_i x_i^{2} w_i}}
\end{pmatrix}
=\mathbf{D}\cdot
\underbrace{\begin{pmatrix}
\dfrac{1}{\sqrt{\sum_i w_i}} & 0\\[10pt]
0 & \dfrac{1}{\sqrt{\sum_i x_i^{2} w_i}}
\end{pmatrix}}_{\mathbf{S}}
=\mathbf{D}\mathbf{S}.
\]

Therefore,
\[
\mathbf{D}_{ul}^{t}\mathbf{D}_{ul}
=\mathbf{S}\,\mathbf{D}^{t}\mathbf{D}\,\mathbf{S}
=\mathbf{S}\,\mathbf{X}^{t}\mathbf{W}\mathbf{X}\,\mathbf{S}
=I(\hat{\boldsymbol{\beta}})_{WS}
=\begin{pmatrix}
1 & \dfrac{\sum_i w_i x_i}{\sqrt{\sum_i w_i}\,\sqrt{\sum_i x_i^{2} w_i}}\\[14pt]
\dfrac{\sum_i w_i x_i}{\sqrt{\sum_i w_i}\,\sqrt{\sum_i x_i^{2} w_i}} & 1
\end{pmatrix},
\]
which coincides with matrix \eqref{eq:info-escalada} and with what \cite{weissfeld1991multicollinearity} proposed.

\subsection*{Appendix C}\label{appendixc}

To assess how often orthogonality is lost in practice, we carried out a simulation study covering the 23 distribution--link combinations in Table~2. For each combination, we considered sample sizes $n \in \{20, 50, 200\}$ and $p \in \{2, 4, 8\}$ explanatory variables in addition to the intercept. In each replication, an exactly orthogonal design $\mathbf{Z} = [\mathbf{1}, \mathbf{X}]$ was generated, with $\mathbf{X}'\mathbf{1} = \mathbf{0}$ and $\mathbf{X}'\mathbf{X}$ diagonal. The GLM was fitted by IRLS and, from the weights at convergence, we computed $CN_{WS}$ and whether $\mathbf{X}'\hat{\mathbf{W}}\mathbf{X}$ is diagonal. Each scenario was replicated 200 times; replications in which the IRLS algorithm did not converge to admissible fitted means (mainly with identity links and $n = 20$) were discarded. Results are presented in Tables \ref{tab:sim_ws} and \ref{tab:sim_mp}.

Both condition numbers show the same pattern. Orthogonality is preserved only in the four combinations with constant weights.   In the remaining 19 combinations, $\mathbf{X}'\hat{\mathbf{W}}\mathbf{X}$ was not diagonal in any replicate. The loss of orthogonality decreases with $n$, but does not disappear, and increases with $p$, particularly for small samples. In most of the scenarios considered here, the differences would not change the conclusion reached with the usual thresholds of 10 and 30. There is, however, one exception: for the exponential model with identity link, $n = 20$ and $p = 8$, the median $CN_{MP}$ (32.54) exceeds the threshold of 30 whereas the median $CN_{WS}$ (28.85) does not, even though the design is orthogonal in $\mathbf{X}'\mathbf{X}$.

\begin{table}[htbp]
\centering
\caption{Simulation results: median and, in parentheses, 95th percentile of $CN_{WS}$, the condition number of \cite{weissfeld1991multicollinearity}, for exactly orthogonal designs ($CN(\mathbf{X}_{ul}'\mathbf{X}_{ul}) = 1$), by sample size $n$ and number of explanatory variables $p$ (200 replications per scenario).}
\label{tab:sim_ws}
\footnotesize
\setlength{\tabcolsep}{2.5pt}
\renewcommand{\arraystretch}{0.95}
\resizebox{\textwidth}{!}{%
\begin{tabular}{ll*{9}{c}}
\toprule
 & & \multicolumn{3}{c}{$n = 20$} & \multicolumn{3}{c}{$n = 50$} & \multicolumn{3}{c}{$n = 200$} \\
\cmidrule(lr){3-5} \cmidrule(lr){6-8} \cmidrule(lr){9-11}
Distribution & Link & $p=2$ & $p=4$ & $p=8$ & $p=2$ & $p=4$ & $p=8$ & $p=2$ & $p=4$ & $p=8$ \\
\midrule
Normal            & Identity     & 1.00 & 1.00 & 1.00 & 1.00 & 1.00 & 1.00 & 1.00 & 1.00 & 1.00 \\
\addlinespace
Exponential       & Inverse      & 1.93 & 2.67 & 5.16 & 1.72 & 1.97 & 2.59 & 1.50 & 1.55 & 1.74 \\
                  &              & {\scriptsize(3.94)} & {\scriptsize(7.04)} & {\scriptsize(13.2)} & {\scriptsize(2.64)} & {\scriptsize(3.29)} & {\scriptsize(4.83)} & {\scriptsize(1.88)} & {\scriptsize(2.07)} & {\scriptsize(2.33)} \\
                  & Log          & 1.00 & 1.00 & 1.00 & 1.00 & 1.00 & 1.00 & 1.00 & 1.00 & 1.00 \\
                  & Identity     & 2.54 & 8.90 & 28.85 & 1.80 & 3.34 & 16.23 & 1.49 & 1.64 & 2.17 \\
                  &              & {\scriptsize(20.0)} & {\scriptsize(146)} & {\scriptsize(391)} & {\scriptsize(14.7)} & {\scriptsize(35.2)} & {\scriptsize(301)} & {\scriptsize(2.10)} & {\scriptsize(7.83)} & {\scriptsize(64.9)} \\
\addlinespace
Gamma             & Inverse      & 1.73 & 1.96 & 2.43 & 1.59 & 1.67 & 1.89 & 1.44 & 1.46 & 1.51 \\
                  &              & {\scriptsize(2.52)} & {\scriptsize(2.97)} & {\scriptsize(3.78)} & {\scriptsize(2.06)} & {\scriptsize(2.19)} & {\scriptsize(2.47)} & {\scriptsize(1.60)} & {\scriptsize(1.68)} & {\scriptsize(1.72)} \\
                  & Log          & 1.00 & 1.00 & 1.00 & 1.00 & 1.00 & 1.00 & 1.00 & 1.00 & 1.00 \\
                  & Identity     & 1.83 & 2.05 & 3.09 & 1.57 & 1.68 & 2.04 & 1.46 & 1.46 & 1.54 \\
                  &              & {\scriptsize(2.83)} & {\scriptsize(3.73)} & {\scriptsize(6.13)} & {\scriptsize(2.07)} & {\scriptsize(2.37)} & {\scriptsize(3.15)} & {\scriptsize(1.66)} & {\scriptsize(1.69)} & {\scriptsize(1.77)} \\
\addlinespace
Inv. Gaussian     & Inv.-square  & 1.80 & 2.49 & 4.75 & 1.59 & 1.82 & 2.74 & 1.38 & 1.45 & 1.62 \\
                  &              & {\scriptsize(4.73)} & {\scriptsize(5.46)} & {\scriptsize(13.4)} & {\scriptsize(2.77)} & {\scriptsize(3.50)} & {\scriptsize(7.05)} & {\scriptsize(1.68)} & {\scriptsize(1.87)} & {\scriptsize(2.13)} \\
                  & Inverse      & 1.32 & 1.46 & 1.78 & 1.25 & 1.32 & 1.45 & 1.20 & 1.22 & 1.26 \\
                  &              & {\scriptsize(1.86)} & {\scriptsize(2.14)} & {\scriptsize(2.53)} & {\scriptsize(1.55)} & {\scriptsize(1.59)} & {\scriptsize(1.92)} & {\scriptsize(1.31)} & {\scriptsize(1.34)} & {\scriptsize(1.37)} \\
                  & Log          & 1.65 & 1.83 & 2.23 & 1.50 & 1.56 & 1.67 & 1.40 & 1.42 & 1.44 \\
                  &              & {\scriptsize(2.10)} & {\scriptsize(2.43)} & {\scriptsize(3.08)} & {\scriptsize(1.74)} & {\scriptsize(1.81)} & {\scriptsize(2.10)} & {\scriptsize(1.52)} & {\scriptsize(1.54)} & {\scriptsize(1.59)} \\
\addlinespace
Poisson           & Log          & 1.64 & 1.69 & 1.98 & 1.50 & 1.54 & 1.61 & 1.40 & 1.41 & 1.44 \\
                  &              & {\scriptsize(1.93)} & {\scriptsize(2.06)} & {\scriptsize(2.52)} & {\scriptsize(1.70)} & {\scriptsize(1.76)} & {\scriptsize(1.85)} & {\scriptsize(1.50)} & {\scriptsize(1.54)} & {\scriptsize(1.54)} \\
                  & Identity     & 1.33 & 1.41 & 1.68 & 1.27 & 1.29 & 1.40 & 1.20 & 1.22 & 1.23 \\
                  &              & {\scriptsize(1.77)} & {\scriptsize(1.90)} & {\scriptsize(2.48)} & {\scriptsize(1.47)} & {\scriptsize(1.52)} & {\scriptsize(1.85)} & {\scriptsize(1.29)} & {\scriptsize(1.30)} & {\scriptsize(1.32)} \\
                  & Square-root  & 1.00 & 1.00 & 1.00 & 1.00 & 1.00 & 1.00 & 1.00 & 1.00 & 1.00 \\
\addlinespace
Binomial          & Logit        & 1.04 & 1.06 & 1.11 & 1.02 & 1.04 & 1.06 & 1.01 & 1.02 & 1.03 \\
                  &              & {\scriptsize(1.11)} & {\scriptsize(1.13)} & {\scriptsize(1.27)} & {\scriptsize(1.06)} & {\scriptsize(1.07)} & {\scriptsize(1.11)} & {\scriptsize(1.03)} & {\scriptsize(1.04)} & {\scriptsize(1.05)} \\
                  & Probit       & 1.05 & 1.09 & 1.15 & 1.03 & 1.05 & 1.08 & 1.02 & 1.03 & 1.04 \\
                  &              & {\scriptsize(1.12)} & {\scriptsize(1.18)} & {\scriptsize(1.32)} & {\scriptsize(1.07)} & {\scriptsize(1.09)} & {\scriptsize(1.13)} & {\scriptsize(1.05)} & {\scriptsize(1.06)} & {\scriptsize(1.06)} \\
                  & Cloglog      & 1.29 & 1.32 & 1.42 & 1.25 & 1.27 & 1.29 & 1.21 & 1.22 & 1.22 \\
                  &              & {\scriptsize(1.44)} & {\scriptsize(1.49)} & {\scriptsize(1.70)} & {\scriptsize(1.34)} & {\scriptsize(1.35)} & {\scriptsize(1.38)} & {\scriptsize(1.26)} & {\scriptsize(1.27)} & {\scriptsize(1.27)} \\
                  & Log          & 1.48 & 1.56 & 1.77 & 1.36 & 1.39 & 1.48 & 1.29 & 1.30 & 1.31 \\
                  &              & {\scriptsize(1.89)} & {\scriptsize(1.85)} & {\scriptsize(2.28)} & {\scriptsize(1.56)} & {\scriptsize(1.59)} & {\scriptsize(1.71)} & {\scriptsize(1.39)} & {\scriptsize(1.39)} & {\scriptsize(1.42)} \\
\addlinespace
Neg. binomial     & Log          & 1.16 & 1.23 & 1.35 & 1.12 & 1.14 & 1.19 & 1.10 & 1.11 & 1.12 \\
                  &              & {\scriptsize(1.36)} & {\scriptsize(1.45)} & {\scriptsize(1.90)} & {\scriptsize(1.22)} & {\scriptsize(1.22)} & {\scriptsize(1.31)} & {\scriptsize(1.15)} & {\scriptsize(1.14)} & {\scriptsize(1.17)} \\
                  & Identity     & 1.69 & 2.47 & 3.60 & 1.52 & 1.79 & 2.53 & 1.36 & 1.44 & 1.56 \\
                  &              & {\scriptsize(3.91)} & {\scriptsize(6641)} & {\scriptsize(6.45)} & {\scriptsize(2.71)} & {\scriptsize(217)} & {\scriptsize(37323)} & {\scriptsize(1.63)} & {\scriptsize(1.81)} & {\scriptsize(2.07)} \\
                  & Canonical    & 1.79 & 2.29 & 3.57 & 1.63 & 1.78 & 2.13 & 1.45 & 1.51 & 1.61 \\
                  &              & {\scriptsize(3.16)} & {\scriptsize(4.13)} & {\scriptsize(7.62)} & {\scriptsize(2.40)} & {\scriptsize(2.72)} & {\scriptsize(3.47)} & {\scriptsize(1.76)} & {\scriptsize(1.85)} & {\scriptsize(1.94)} \\
\addlinespace
Bell              & Log          & 1.46 & 1.58 & 1.83 & 1.37 & 1.39 & 1.48 & 1.29 & 1.29 & 1.33 \\
                  &              & {\scriptsize(1.79)} & {\scriptsize(1.95)} & {\scriptsize(2.52)} & {\scriptsize(1.58)} & {\scriptsize(1.60)} & {\scriptsize(1.70)} & {\scriptsize(1.39)} & {\scriptsize(1.40)} & {\scriptsize(1.43)} \\
                  & Identity     & 1.50 & 1.74 & 2.29 & 1.36 & 1.49 & 1.77 & 1.26 & 1.31 & 1.35 \\
                  &              & {\scriptsize(2.32)} & {\scriptsize(2.56)} & {\scriptsize(9918)} & {\scriptsize(1.84)} & {\scriptsize(2.54)} & {\scriptsize(1659)} & {\scriptsize(1.42)} & {\scriptsize(1.48)} & {\scriptsize(1.64)} \\
                  & Square-root  & 1.14 & 1.18 & 1.27 & 1.11 & 1.13 & 1.16 & 1.09 & 1.10 & 1.10 \\
                  &              & {\scriptsize(1.26)} & {\scriptsize(1.33)} & {\scriptsize(1.50)} & {\scriptsize(1.20)} & {\scriptsize(1.21)} & {\scriptsize(1.27)} & {\scriptsize(1.12)} & {\scriptsize(1.13)} & {\scriptsize(1.13)} \\
\bottomrule
\end{tabular}
}
\end{table}

\begin{table}[htbp]
\centering
\caption{Simulation results: median and, in parentheses, 95th percentile of $CN_{MP}$, the condition number of \cite{mackinnon1989collinearity}, for the same designs and replications as in Table~\ref{tab:sim_ws}.}
\label{tab:sim_mp}
\footnotesize
\setlength{\tabcolsep}{2.5pt}
\resizebox{\textwidth}{!}{%
\begin{tabular}{ll*{9}{c}}
\toprule
 & & \multicolumn{3}{c}{$n = 20$} & \multicolumn{3}{c}{$n = 50$} & \multicolumn{3}{c}{$n = 200$} \\
\cmidrule(lr){3-5} \cmidrule(lr){6-8} \cmidrule(lr){9-11}
Distribution & Link & $p=2$ & $p=4$ & $p=8$ & $p=2$ & $p=4$ & $p=8$ & $p=2$ & $p=4$ & $p=8$ \\
\midrule
Normal            & Identity     & 1.00 & 1.00 & 1.00 & 1.00 & 1.00 & 1.00 & 1.00 & 1.00 & 1.00 \\
\addlinespace
Exponential       & Inverse      & 2.00 & 2.96 & 5.41 & 1.75 & 2.05 & 2.74 & 1.53 & 1.58 & 1.81 \\
                  &              & {\scriptsize(4.21)} & {\scriptsize(6.67)} & {\scriptsize(14.0)} & {\scriptsize(2.92)} & {\scriptsize(3.65)} & {\scriptsize(5.01)} & {\scriptsize(2.00)} & {\scriptsize(2.17)} & {\scriptsize(2.43)} \\
                  & Log          & 1.00 & 1.00 & 1.00 & 1.00 & 1.00 & 1.00 & 1.00 & 1.00 & 1.00 \\
                  & Identity     & 2.68 & 9.65 & 32.54 & 1.85 & 3.72 & 17.44 & 1.52 & 1.67 & 2.30 \\
                  &              & {\scriptsize(22.2)} & {\scriptsize(159)} & {\scriptsize(356)} & {\scriptsize(17.0)} & {\scriptsize(42.4)} & {\scriptsize(308)} & {\scriptsize(2.25)} & {\scriptsize(8.82)} & {\scriptsize(53.3)} \\
\addlinespace
Gamma             & Inverse      & 1.79 & 2.08 & 2.65 & 1.63 & 1.71 & 1.94 & 1.46 & 1.48 & 1.52 \\
                  &              & {\scriptsize(2.77)} & {\scriptsize(3.23)} & {\scriptsize(4.21)} & {\scriptsize(2.18)} & {\scriptsize(2.30)} & {\scriptsize(2.54)} & {\scriptsize(1.64)} & {\scriptsize(1.71)} & {\scriptsize(1.76)} \\
                  & Log          & 1.00 & 1.00 & 1.00 & 1.00 & 1.00 & 1.00 & 1.00 & 1.00 & 1.00 \\
                  & Identity     & 1.88 & 2.15 & 3.31 & 1.60 & 1.70 & 2.11 & 1.47 & 1.48 & 1.57 \\
                  &              & {\scriptsize(3.38)} & {\scriptsize(3.74)} & {\scriptsize(6.74)} & {\scriptsize(2.29)} & {\scriptsize(2.50)} & {\scriptsize(3.53)} & {\scriptsize(1.69)} & {\scriptsize(1.74)} & {\scriptsize(1.83)} \\
\addlinespace
Inv. Gaussian     & Inv.-square  & 1.82 & 2.62 & 5.20 & 1.63 & 1.88 & 2.86 & 1.39 & 1.47 & 1.65 \\
                  &              & {\scriptsize(5.10)} & {\scriptsize(6.13)} & {\scriptsize(14.4)} & {\scriptsize(3.23)} & {\scriptsize(3.87)} & {\scriptsize(7.18)} & {\scriptsize(1.76)} & {\scriptsize(2.04)} & {\scriptsize(2.23)} \\
                  & Inverse      & 1.34 & 1.50 & 1.83 & 1.26 & 1.34 & 1.46 & 1.21 & 1.22 & 1.27 \\
                  &              & {\scriptsize(1.91)} & {\scriptsize(2.18)} & {\scriptsize(2.68)} & {\scriptsize(1.63)} & {\scriptsize(1.62)} & {\scriptsize(1.99)} & {\scriptsize(1.32)} & {\scriptsize(1.36)} & {\scriptsize(1.38)} \\
                  & Log          & 1.70 & 1.87 & 2.30 & 1.52 & 1.58 & 1.70 & 1.41 & 1.42 & 1.45 \\
                  &              & {\scriptsize(2.15)} & {\scriptsize(2.52)} & {\scriptsize(3.22)} & {\scriptsize(1.77)} & {\scriptsize(1.87)} & {\scriptsize(2.12)} & {\scriptsize(1.54)} & {\scriptsize(1.56)} & {\scriptsize(1.60)} \\
\addlinespace
Poisson           & Log          & 1.68 & 1.73 & 2.08 & 1.52 & 1.55 & 1.63 & 1.41 & 1.42 & 1.45 \\
                  &              & {\scriptsize(1.99)} & {\scriptsize(2.14)} & {\scriptsize(2.60)} & {\scriptsize(1.71)} & {\scriptsize(1.78)} & {\scriptsize(1.86)} & {\scriptsize(1.51)} & {\scriptsize(1.55)} & {\scriptsize(1.55)} \\
                  & Identity     & 1.34 & 1.45 & 1.74 & 1.28 & 1.31 & 1.41 & 1.20 & 1.22 & 1.24 \\
                  &              & {\scriptsize(1.81)} & {\scriptsize(2.07)} & {\scriptsize(2.52)} & {\scriptsize(1.51)} & {\scriptsize(1.57)} & {\scriptsize(1.88)} & {\scriptsize(1.30)} & {\scriptsize(1.31)} & {\scriptsize(1.33)} \\
                  & Square-root  & 1.00 & 1.00 & 1.00 & 1.00 & 1.00 & 1.00 & 1.00 & 1.00 & 1.00 \\
\addlinespace
Binomial          & Logit        & 1.06 & 1.08 & 1.12 & 1.05 & 1.05 & 1.07 & 1.03 & 1.03 & 1.04 \\
                  &              & {\scriptsize(1.15)} & {\scriptsize(1.17)} & {\scriptsize(1.27)} & {\scriptsize(1.07)} & {\scriptsize(1.09)} & {\scriptsize(1.11)} & {\scriptsize(1.04)} & {\scriptsize(1.04)} & {\scriptsize(1.05)} \\
                  & Probit       & 1.09 & 1.11 & 1.17 & 1.06 & 1.07 & 1.10 & 1.04 & 1.05 & 1.05 \\
                  &              & {\scriptsize(1.15)} & {\scriptsize(1.20)} & {\scriptsize(1.33)} & {\scriptsize(1.09)} & {\scriptsize(1.11)} & {\scriptsize(1.15)} & {\scriptsize(1.06)} & {\scriptsize(1.06)} & {\scriptsize(1.07)} \\
                  & Cloglog      & 1.30 & 1.34 & 1.44 & 1.25 & 1.27 & 1.29 & 1.21 & 1.22 & 1.22 \\
                  &              & {\scriptsize(1.47)} & {\scriptsize(1.50)} & {\scriptsize(1.71)} & {\scriptsize(1.35)} & {\scriptsize(1.36)} & {\scriptsize(1.39)} & {\scriptsize(1.26)} & {\scriptsize(1.27)} & {\scriptsize(1.27)} \\
                  & Log          & 1.49 & 1.61 & 1.83 & 1.38 & 1.41 & 1.51 & 1.29 & 1.30 & 1.32 \\
                  &              & {\scriptsize(1.94)} & {\scriptsize(1.94)} & {\scriptsize(2.44)} & {\scriptsize(1.57)} & {\scriptsize(1.64)} & {\scriptsize(1.77)} & {\scriptsize(1.40)} & {\scriptsize(1.40)} & {\scriptsize(1.42)} \\
\addlinespace
Neg. binomial     & Log          & 1.17 & 1.25 & 1.38 & 1.12 & 1.14 & 1.20 & 1.10 & 1.11 & 1.12 \\
                  &              & {\scriptsize(1.38)} & {\scriptsize(1.47)} & {\scriptsize(1.97)} & {\scriptsize(1.23)} & {\scriptsize(1.22)} & {\scriptsize(1.32)} & {\scriptsize(1.15)} & {\scriptsize(1.14)} & {\scriptsize(1.17)} \\
                  & Identity     & 1.78 & 2.57 & 3.77 & 1.56 & 1.86 & 2.65 & 1.37 & 1.46 & 1.58 \\
                  &              & {\scriptsize(4.29)} & {\scriptsize(5915)} & {\scriptsize(6.87)} & {\scriptsize(2.97)} & {\scriptsize(213)} & {\scriptsize(36121)} & {\scriptsize(1.67)} & {\scriptsize(1.93)} & {\scriptsize(2.20)} \\
                  & Canonical    & 1.86 & 2.47 & 3.80 & 1.65 & 1.84 & 2.22 & 1.47 & 1.52 & 1.64 \\
                  &              & {\scriptsize(3.38)} & {\scriptsize(4.72)} & {\scriptsize(8.09)} & {\scriptsize(2.53)} & {\scriptsize(2.86)} & {\scriptsize(3.77)} & {\scriptsize(1.83)} & {\scriptsize(1.92)} & {\scriptsize(2.04)} \\
\addlinespace
Bell              & Log          & 1.48 & 1.62 & 1.90 & 1.39 & 1.40 & 1.51 & 1.30 & 1.30 & 1.33 \\
                  &              & {\scriptsize(1.85)} & {\scriptsize(2.02)} & {\scriptsize(2.59)} & {\scriptsize(1.59)} & {\scriptsize(1.62)} & {\scriptsize(1.75)} & {\scriptsize(1.39)} & {\scriptsize(1.40)} & {\scriptsize(1.44)} \\
                  & Identity     & 1.52 & 1.80 & 2.35 & 1.38 & 1.52 & 1.82 & 1.27 & 1.32 & 1.36 \\
                  &              & {\scriptsize(2.54)} & {\scriptsize(2.77)} & {\scriptsize(8105)} & {\scriptsize(1.94)} & {\scriptsize(2.70)} & {\scriptsize(1158)} & {\scriptsize(1.45)} & {\scriptsize(1.52)} & {\scriptsize(1.67)} \\
                  & Square-root  & 1.14 & 1.19 & 1.29 & 1.12 & 1.14 & 1.17 & 1.09 & 1.10 & 1.10 \\
                  &              & {\scriptsize(1.27)} & {\scriptsize(1.36)} & {\scriptsize(1.53)} & {\scriptsize(1.20)} & {\scriptsize(1.22)} & {\scriptsize(1.29)} & {\scriptsize(1.12)} & {\scriptsize(1.13)} & {\scriptsize(1.13)} \\
\bottomrule
\end{tabular}}
\end{table}

\newpage
\bibliographystyle{apalike}
\bibliography{bib}

\end{document}